\documentclass[fleqn,usenatbib]{mnras}

\usepackage[T1]{fontenc}

\DeclareRobustCommand{\VAN}[3]{#2}
\let\VANthebibliography\thebibliography
\def\thebibliography{\DeclareRobustCommand{\VAN}[3]{##3}\VANthebibliography}

\usepackage{graphicx}	
\usepackage{amsmath}	
\usepackage{amssymb}	

\usepackage{xspace}        
\usepackage{xcolor}          

\usepackage[normalem]{ulem}

\usepackage{newtxtext,newtxmath}

\definecolor{purple}{RGB}{76, 0,153}

\definecolor{hotpink}{RGB}{255, 0,255}

\definecolor{darkolivegreen}{rgb}{0.33, 0.42, 0.18}

\newcommand{\Bnl}{\beta^\mathrm{NL}}
\newcommand{\nbody}{$N$-body\xspace}
\newcommand{\eg}{e.g.,\xspace}
\newcommand{\ie}{i.e.\xspace}
\newcommand{\sform}[2]{{#1}\times10^{#2}}
\newcommand{\Mpc}{\,h^{-1}\mathrm{Mpc}}
\newcommand{\iMpc}{\,h\mathrm{Mpc}^{-1}}

\newcommand{\Om}{\Omega_\mathrm{m}}
\newcommand{\Ow}{\Omega_w}
\newcommand{\oc}{\omega_\mathrm{c}}
\newcommand{\ob}{\omega_\mathrm{b}}
\newcommand{\ns}{n_\mathrm{s}}
\newcommand{\As}{A_\mathrm{s}}
\newcommand{\DQ}{Dark Quest\xspace}

\title[Halo model cosmology]{The halo model with beyond-linear halo bias: unbiasing cosmological constraints from galaxy-galaxy lensing and clustering}

\author[C. Mahony et al.]{Constance Mahony$^{1}$\thanks{E-mail: mahony@astro.rub.de},
Andrej Dvornik$^{1}$,
Alexander Mead$^{1,2}$,
Catherine Heymans$^{1,2}$,
Marika Asgari$^{2,3}$,
\newauthor
Hendrik Hildebrandt$^{1}$,
Hironao Miyatake$^{4,5}$,
Takahiro Nishimichi$^{5,6}$,
Robert Reischke$^{1}$
\\
$^{1}$Ruhr University Bochum, Faculty of Physics and Astronomy, Astronomical Institute (AIRUB), German Centre for Cosmological Lensing, \\ 44780 Bochum, Germany\\
$^{2}$Institute for Astronomy, University of Edinburgh, Royal Observatory, Blackford Hill, Edinburgh, EH9 3HJ, UK\\
$^{3}$E. A. Milne Centre, University of Hull, Cottingham Road, Hull, HU6 7RX, UK\\
$^{4}$Kobayashi-Maskawa Institute for the Origin of Particles and the Universe (KMI), Nagoya University, Nagoya, 464-8602, Japan\\
$^{5}$Kavli Institute for the Physics and Mathematics of the Universe (WPI), The University of Tokyo Institutes for Advanced Study (UTIAS), \\ The University of Tokyo, Chiba 277-8583, Japan\\
$^{6}$Center for Gravitational Physics, Yukawa Institute for Theoretical Physics, Kyoto University, Kyoto 606-8502, Japan}

\date{Accepted XXX. Received YYY; in original form ZZZ}

\pubyear{2015}

\begin{document}
\label{firstpage}
\pagerange{\pageref{firstpage}--\pageref{lastpage}}
\maketitle

\begin{abstract}
We determine the error introduced in a joint halo model analysis of galaxy-galaxy lensing and galaxy clustering observables when adopting the standard approximation of linear halo bias.   Considering the Kilo-Degree Survey, we forecast that ignoring the non-linear halo bias would result in up to 5$\sigma$ offsets in the recovered cosmological parameters describing structure growth, $S_8$, and the matter density parameter, $\Omega_{\mathrm{m}}$. We include the scales $10^{-1.3}<r_{\rm{p}} \ / h^{-1}\, \mathrm{Mpc}<10$ in the data vector, and the direction of these offsets are shown to depend on the freedom afforded to the halo model through other nuisance parameters.  We conclude that a beyond-linear halo bias correction must therefore be included in future cosmological halo model analyses of large-scale structure observables on non-linear scales.
\end{abstract}




\section{Introduction}

The halo model is a phenomenological model often used to interpret the large-scale structure of the Universe (see \citealt{Cooray:2002dia} for a review). In this model all dark matter exists within dark matter halos, which trace the underlying matter fluctuations. In its most generic form it includes a number of approximations such as dark matter halos are spherical and can be completely described by their mass, and that the halos trace the underlying matter fluctuations in a linearly biased way -- linear halo bias. These assumptions have provided a useful description of large-scale structure observables until now, but with ever improving datasets these need to be revisited. In this paper we focus on the impact of neglecting the non-linear nature of halo bias.

Galaxy-galaxy lensing studies are concerned with matter-galaxy overdensity correlations and often use a halo model to interpret the data, and to understand the connection between galaxies and halo formation (e.g. \citealt{Mandelbaum:2004mq,Cacciato:2008hm})\footnote{Large survey area spectroscopic galaxy clustering surveys tend to cut scales relevant to the halo model \citep{BOSS:2016wmc,eBOSS:2020yzd}.}. It is common to assume linear halo bias in halo models of galaxy-galaxy lensing (e.g. \citealt{Cacciato2012,Dvornik:2018frx,DES:2021olg}), or to include some non-linear halo bias through techniques such as `halo exclusion' where halos are not allowed to overlap \citep{2013BoschCacciato}. In the case of matter-matter correlations it is possible to use fitting functions (e.g. HALOFIT \citealt{Smith:2002dz, Takahashi:2012em}) or phenomenological parameters (e.g. HM-CODE \citealt{HMxCode}) to overcome the limitations of the halo model. However, once galaxy correlations are included these corrections are no longer applicable, as they do not connect the galaxy distribution to non-linear halo bias \citep{Mead:2020qdk}.

\cite{Mead:2020qdk} explore the relation between halos and the underlying matter distribution, and address the standard approximation that halos trace the underlying matter distribution with a linear halo bias. They measure the non-linear halo bias from \textit{N}-body simulations, incorporating an additional beyond-linear halo bias correction, $\beta^{\mathrm{NL}}$, into the halo modelling. A key benefit is that the correction, $\beta^{\mathrm{NL}}$, can easily be included into the existing halo model framework.  \cite{Miyatake:2020uhg} present a complementary approach to account for beyond-linear halo bias, directly emulating the galaxy-galaxy lensing and galaxy clustering observables from similar non-linearly biased simulations \citep{DarkEmulatorCode}. We compare these two approaches in Section~\ref{sec:DEcomp} and find them to be consistent.

In this paper we present forecasts for a joint halo model cosmological analysis of galaxy-galaxy lensing and galaxy clustering observables with the Kilo-Degree Survey \citep[KiDS, ][]{Kuijken2019}.  Section \ref{sec:including bias} details how the \cite{Mead:2020qdk} beyond-linear halo bias correction, $\beta^{\mathrm{NL}}$, is incorporated into the halo model power spectra where linear halo bias has previously been assumed \citep{Dvornik:2018frx}. Section \ref{sec: observables} presents how the lensing and clustering observables are impacted by the beyond-linear halo bias correction.  The resulting cosmological parameter offsets introduced by a standard linear halo model analysis of joint lensing-clustering observables is presented in Section~\ref{sec: cosmological parameters}.  Through an analysis of mock data from the {\sc DarkEmulator} \citep{Nishimichi2019,DarkEmulatorCode}, we demonstrate in Section~\ref{sec:DEcomp} that the accurate recovery of cosmological parameters is possible when incorporating $\beta^{\mathrm{NL}}$ into the halo model. We conclude in Section \ref{sec: conclusions}.  In Appendix~\ref{app:DarkEmu} we present details of the simulations used in this work, and in Appendix~\ref{sec:rescaling} a rescaling technique to model the cosmology dependence of $\beta^{\mathrm{NL}}$. 

\section{Including Beyond-linear Halo Bias} \label{sec:including bias}

The halo model assumes that all dark matter exists within dark matter halos, which are then populated with galaxies. In this Section we summarise how galaxies populate dark matter halos using the conditional stellar mass function (CSMF) formalism \citep{Yang:2007pg,Cacciato:2008hm,Cacciato2013MNRAS.430..767C,Wang2013,vanUitert2016,Dvornik:2018frx}. A key feature of the CSMF formalism is that the galaxies are split into centrals and satellites, where centrals reside at the centre of their host halo and satellites orbit around them.

In this work we require two 3D power spectra to calculate our observables: the galaxy-galaxy power spectrum $P_{\rm{gg}}$, and the galaxy-matter power spectrum $P_{\rm{g\delta}}$. These can be split into contributions from one-halo (1h) and two-halo (2h) terms, where the 1h term describes the clustering on small scales within a single halo and the 2h term describes the clustering on larger scales between two halos. These terms can then be further broken down into contributions from central (c) and satellite (s) galaxies,
\begin{equation}
  \begin{split}
  & P_{\rm{gg}} = 2P_{\rm{cs}}^{\rm{1h}} + P_{\rm{ss}}^{\rm{1h}} + P_{\rm{cc}}^{\rm{2h}} + 2P_{\rm{cs}}^{\rm{2h}} + P_{\rm{ss}}^{\rm{2h}} \ , \\
  & P_{\rm{g\delta}} = P_{\rm{c\delta}}^{\rm{1h}} + P_{\rm{s\delta}}^{\rm{1h}} + P_{\rm{c\delta}}^{\rm{2h}} + P_{\rm{s\delta}}^{\rm{2h}} \ .
 \end{split}
\end{equation}
We do not include $P_{\rm{cc}}^{\rm{1h}}$ as there is only one central galaxy per halo so this term corresponds to shot noise, which we do not include in the measurements. As shown in \cite{2013BoschCacciato,Cacciato2013MNRAS.430..767C,Dvornik:2018frx} these contributions are given by,
\begin{equation}
  \begin{split}
  P_{\rm{xy}}^{\rm{1h}}(k,z) = & \int_0^\infty \mathcal{H}_{\rm{x}}(k, M, z)\mathcal{H}_{\rm{y}}(k,M,z)n(M,z)\mathrm{d}M \ ,\\
  P_{\rm{xy}}^{\rm{2h}}(k,z) = & P^\mathrm{lin}_{\delta \delta} (k,z) \int_0^\infty  \mathrm{d}M_1 \mathcal{H}_{\rm{x}}(k, M_1,z)n(M_1,z) b(M_1,z) \\
  & \times \int_0^\infty \mathrm{d}M_2 \mathcal{H}_{\rm{y}}(k,M_2,z)n(M_2,z) b(M_2,z) \ ,
 \end{split}\label{eq:standard 1h 2h}
\end{equation}
where x and y can be c, s or $\delta$. $P^\mathrm{lin}_{\delta \delta}$ is the linear matter power spectrum, which we obtain using the \cite{Eisenstein:1997ik} transfer function. We calibrate the halo mass function, $n(M,z)$, the number density of dark matter halos with mass $M$ at redshift $z$, and the halo bias, $b(M,z)$, which accounts for dark matter halos being linearly biased tracers of the underlying dark matter distribution, from numerical simulations \citep{2010tinker}. The profiles $\mathcal{H}$ encode the matter or galaxy contribution,
\begin{equation}\label{eq:Hfactors}
  \begin{split}
  & \mathcal{H}_{\rm{\delta}}(k,M,z) = \frac{M}{\bar{\rho}_{\rm{m}}}\tilde{u}_{\rm{h}}(k|M,z) \ , \\
  & \mathcal{H}_{\rm{c}}(k,M,z)=\mathcal{H}_{\rm{c}}(M,z) = \frac{\langle N_{\rm{c}}|M \rangle}{\bar{n}_{\rm{c}}(z)} \ , \\
  &\mathcal{H}_{\rm{s}}(k,M,z) = \frac{\langle N_{\rm{s}} | M \rangle}{\bar{n}_{\rm{s}}(z)}\tilde{u}_{\rm{s}}(k|M,z) \ ,
  \end{split}
\end{equation}
where $\bar{\rho}_m$ is the present day mean matter density of the Universe. 

The average number of central and satellite galaxies in a halo of mass $M$ within the stellar mass range [$M_{\star,1}$, $M_{\star,2}$], $\langle N_{\rm{c}}|M\rangle$ and $\langle N_{\rm{s}}|M\rangle$, and the average number density of central and satellite galaxies across all halo masses, $\bar{n}_{\rm{c}}$ and $\bar{n}_{\rm{s}}$, are the Halo Occupation Distribution (HOD) quantities. These are computed using the CSMF formalism,
\begin{equation}\label{HODfromCLF}
  \langle N_{\mathrm{x}} \vert M \rangle = \int_{M_{\star,1}}^{M_{\star,2}} \Phi_{\mathrm{x}}(M_{\star}  \vert M) \, \mathrm{d} M_{\star} \ ,
  \end{equation}
and,
  \begin{equation}\label{averng}
    \overline{n}_{\mathrm{x}} = \int_{0}^{\infty} \langle N_{\mathrm{x}} \vert M \rangle \, n(M) \, \mathrm{d} M\ .
    \end{equation}
Here $\Phi_{\mathrm{x}}(M_{\star}  \vert M)$ denotes the CSMF, the  average  number  of  galaxies with stellar mass $M_{\star}$ that reside in a halo of mass $M$. Note $\langle N_{\rm{c}}|M\rangle$ varies between 0 and 1, as there is at most one central galaxy per halo. The CSMF of central galaxies is modelled as a log-normal,
\begin{equation}\label{phi_c}
  \Phi_{\mathrm{c}}(M_{\star}  \vert M) = {1 \over {\sqrt{2\pi} \, {\ln}(10)\, \sigma_{\mathrm{c}} M_{\star} } 
  }{\exp}\left[- { {\log(M_{\star} / M^{*}_{\mathrm{c}} )^2 } \over 2\,\sigma_{\mathrm{c}}^{2}} \right]\, \,,
  \end{equation}
where $\sigma_{\mathrm{c}}$ is the scatter between stellar mass and halo mass and $M^{*}_{\mathrm{c}}$ is parameterised as, 
\begin{equation}\label{eq:CMF4}
  M^{*}_{\mathrm{c}}(M) = M_{0} \frac{(M/M_{1})^{\gamma_{1}}}{[1 + (M/M_{1})]^{\gamma_{1} - \gamma_{2}}} \ ,
  \end{equation}
where $M_0$, $M_1$, $\gamma_{1}$ and $\gamma_{2}$ are free parameters. The CSMF of satellite galaxies is modelled as a modified Schechter function,
\begin{equation}\label{phi_s}
  \Phi_{\mathrm{s}}(M_{\star}  \vert M) = { \phi^{*}_{\mathrm{s}} \over M^{*}_{\mathrm{s}}}\,
  \left({M_{\star} \over M^{*}_{\mathrm{s}}}\right)^{\alpha_{\mathrm{s}}} \,
  {\exp} \left[- \left ({M_{\star} \over M^{*}_{\mathrm{s}}}\right )^2 \right] 
  \,,
  \end{equation}
where $\alpha_{\mathrm{s}}$ governs the power law behaviour of satellite galaxies, $M_{\mathrm{s}}^{*}$ is parametrised as, 
  \begin{equation}\label{eq:CMF5}
    M_{\mathrm{s}}^{*}(M) = 0.56\ M^{*}_{\mathrm{c}}(M)\,,
    \end{equation}
    and $\phi_{\mathrm{s}}^{*}$ is parametrised as,
    \begin{equation}\label{eq:CMF7}
    \log[\phi_{\mathrm{s}}^{*}(M)] = b_{1} + b_{2}(\log m_{13})\,,
    \end{equation}
where $m_{13} = M/(10^{13}M_{\odot}h^{-1})$, and $b_{1}$ and $b_{2}$ are free parameters. These parameterisations are motivated by \citet{Yang:2007pg}. For further details of the CSMF formalism see \citet{Cacciato2013MNRAS.430..767C} and \citet{Dvornik:2018frx}.

Referring back to equation \ref{eq:Hfactors}, $\tilde{u}_{\rm{h}}$ is the Fourier transform of the normalised density distribution of dark matter in a halo of mass $M$, and $\tilde{u}_{\rm{s}}$ is the normalised number density distribution of satellite galaxies in a halo of mass $M$. There is no $\tilde{u}_{\rm{c}}$ as there is only one central galaxy per halo. We assume satellites follow the spatial distribution of the underlying dark matter, i.e. $\tilde{u}_{\rm{s}} \equiv \tilde{u}_{\rm{h}}$, and assume that the density profile of dark matter haloes follows an NFW profile \citep{1996NFWN}. The NFW profile is described by two parameters the concentration, $c$, and mass, $M$, of the halo, however these two parameters are correlated. In this work we adopt the \cite{Duffy_2008} concentration-mass relation,
\begin{equation}
    \label{eq:con_duffy}
    c(M, z) = 10.14\; \ \left[\frac{M}{(2\times 10^{12} M_{\odot}/h)}\right]^{- 0.081}\ (1+z)^{-1.01} \ ,
\end{equation}
and additionally include two normalisations,
\begin{equation}\label{eq:f_h,s}
  c_{\mathrm{h,s}}(M, z) = f_{\mathrm{h,s}}\, c(M, z)\,,
\end{equation} 
where $f_{\mathrm{h}}$ normalises the concentration-mass relation for the distribution of dark matter $\tilde{u}_{\rm{h}}$ and $f_{\mathrm{s}}$ normalises the concentration-mass relation for the distribution of satellite galaxies $\tilde{u}_{\rm{s}}$. \cite{Debackere:2021ado} show that including these parameters can help to account for the impact of baryonic feedback.

The two-halo term in equation \ref{eq:standard 1h 2h} assumes that haloes are linearly biased tracers of the underlying matter field,
\begin{equation}
  P_{\rm{hh}}(M_1,M_2,k,z) \simeq b(M_1,z)b(M_2,z)P^\mathrm{lin}_{\delta \delta} (k,z).
\end{equation}
\cite{Mead:2020qdk} address this assumption by introducing a beyond-linear bias correction $\beta^{\mathrm{NL}}$ so,  
\begin{equation}
  \begin{split}
  P_{\rm{hh}}&(M_1,M_2,k,z) \simeq \\
  & b(M_1,z)b(M_2,z)P^\mathrm{lin}_{\delta \delta} (k,z)[1+\beta^{\mathrm{NL}}(M_1,M_2,k,z)], 
  \label{eq:bnl hh}
\end{split}
\end{equation}
and the $\beta^{\mathrm{NL}}$ function encompasses everything beyond the linear bias model. The 2h terms in equation \ref{eq:standard 1h 2h} then become,
\begin{equation}
\begin{split}
  P_{\rm{xy}}^{\rm{2h}}(k,z) & = P^\mathrm{lin}_{\delta \delta} (k,z) \int_0^\infty  \mathrm{d}M_1 \mathcal{H}_{\rm{x}}(k, M_1,z)n(M_1,z) b(M_1,z) \\
  & \times \int_0^\infty  \mathrm{d}M_2 \mathcal{H}_{\rm{y}}(k,M_2,z)n(M_2,z) b(M_2,z) \ \\
  & + P^\mathrm{lin}_{\delta \delta} (k,z) I_{\mathrm{xy}}^{\mathrm{NL}}(k,z) \,
 \end{split}
 \label{eq:2 halo term}
\end{equation}
where the second term includes the \cite{Mead:2020qdk} beyond-linear halo bias correction $\beta^{\mathrm{NL}}$,
\begin{equation}
\begin{split}
I_{\mathrm{xy}}^{\mathrm{NL}}(k,z) & = \int_0^\infty  \int_0^\infty  \mathrm{d}M_1 \mathrm{d}M_2 \ \beta^{\mathrm{NL}}(k,M_1,M_2,z) \\
& \times \mathcal{H}_{\rm{x}}(k, M_1,z)\mathcal{H}_{\rm{y}}(k,M_2,z)n(M_1,z)\\
& \times n(M_2,z)b(M_1,z)b(M_2,z) \ . 
\end{split}
\end{equation}
$\beta^{\mathrm{NL}}$ is calibrated directly from numerical simulations using simulation-measured quantities of the linear bias $\hat{b}$ on large scales, and the halo auto power spectrum $\hat{P}_\mathrm{hh}$. In the large-scale limit $\beta^\mathrm{NL}(M_1,M_2,k \rightarrow 0, z) = 0$, such that equation~\ref{eq:2 halo term} returns to the standard linear halo model formalism with $I_{\mathrm{xy}}^{\mathrm{NL}}(k \rightarrow 0,z)=0$.

In this work we extend the \cite{Mead:2020qdk} analysis by calibrating $\beta^{\mathrm{NL}}$ for a range of different cosmologies utilising the Dark Quest $N$-body simulations \citep{Nishimichi2019,DarkEmulatorCode}, instead of \textsc{Multidark}\footnote{\textsc{Multidark}:\url{https://www.cosmosim.org/}} \citep{Klypin2011,Prada2012,Riebe2013AN....334..691R}. Dark Quest explores a six-dimensional cosmological parameter space within the $w$CDM framework sampled with $100$ models following a sliced latin hypercube design, centered at the \cite{Planck2020} best-fit cosmological model. The partner \textsc{DarkEmulator}\footnote{\textsc{DarkEmulator}:\url{https://github.com/DarkQuestCosmology/dark_emulator_public}} regressor utilises Gaussian Processes and a weighted Principal Component Analysis to then make predictions for quantities measured by Dark Quest, for any set of cosmological parameters within the support range of the training simulations \citep{DarkEmulatorCode}. See Appendix~\ref{app:DarkEmu} for further details.

\section{Observables} \label{sec: observables}

The two observables included in this analysis are the projected galaxy-galaxy correlation function $w_{\rm{p}}(r_{\rm{p}},z)$ (galaxy clustering) and the excess surface density profile $\Delta\Sigma(r_{\rm{p}},z)$ (galaxy-galaxy lensing). These are calculated from the power-spectra $P_{\rm{gg}}$ and $P_{\rm{g\delta}}$ by computing the two-point correlation functions,
\begin{equation}
  \xi_{\rm{gx}}(r,z) = \frac{1}{2\pi^2}\int_0^\infty P_{\rm{gx}}(k,z)\frac{{\rm{sin}} \ kr}{kr}k^2 \ \mathrm{d} k \ ,
\end{equation}
where x is either g or $\delta$.  The projected galaxy-galaxy correlation function $w_{\rm{p}}(r_{\rm{p}},z)$ relates to the 3D galaxy-galaxy correlation function $\xi_{\rm{gg}}(r,z)$ via,
\begin{equation}
  w_{\rm{p}}(r_{\rm{p}},z) = 2\int_0^{r_{\pi,\mathrm{max}}} \xi_{\rm{gg}}(r_{\mathrm{p}},r_{\pi}, z) \ {\rm{d}}r_{\pi} \ ,
\end{equation}
where $r_{\rm{p}}$ is the projected separation between two galaxies, $r_{\pi}$ the separation perpendicular to the line-of-sight and $r_{\pi,\mathrm{max}}$ the maximum integration range used for the data ($r_{\pi,\mathrm{max}} = 100 \ \mathrm{Mpc}/h$ in this work). The excess surface density profile $\Delta\Sigma(r_{\rm{p}},z)$ is given by,
\begin{equation}
  \Delta\Sigma(r_{\rm{p}},z) = \frac{2}{r_{\rm{p}}^2}\int^{r_{\rm{p}}}_0 \Sigma(R',z)R'{\rm{d}} R' - \Sigma(r_{\rm{p}},z) \ ,
\end{equation}
where $\Sigma(r_{\rm{p}},z)$ is the projected surface mass density. This relates to the galaxy-matter correlation function $\xi_{\rm{g\delta}}(r,z)$ via,
\begin{equation}
  \Sigma(r_{\rm{p}},z)=2\bar{\rho}_{\rm{m}} \int_{r_{\rm{p}}}^\infty\xi_{\rm{g\delta}}(r,z)\frac{r{\rm{d}}r}{\sqrt{r^2-r_{\rm{p}}^2}} ,
\end{equation}
where $\bar{\rho}_{\rm{m}}$ is the present day mean matter density of the universe. For further details see \cite{Dekel:1998eq,Sheldon2004,Cacciato2013MNRAS.430..767C}; and \cite{Dvornik:2018frx}.

In this work we forecast the impact of including a beyond-linear halo bias correction $\beta^{\mathrm{NL}}$ for a galaxy clustering sample similar to the bright galaxy sample in the Kilo-Degree Survey Data Release 4 \citep{Bilicki:2021hgn}. This sample is flux-limited at $r$ < 20 mag and contains approximately 1 million galaxies with a mean redshift of 0.23. It has a similar selection to the Galaxy And Mass Assembly survey (GAMA, \citealt{driver2011MNRAS.413..971D}). We therefore use the CSMF parameter values found for GAMA galaxies in \cite{vanUitert2016} to simulate the power spectra, and hence the observables. Referring to section \ref{sec:including bias}, there are 10 CSMF parameters $[f_{\mathrm{h}}, M_0, M_1, \gamma_1, \gamma_2, \sigma_c, f_{\mathrm{s}}, \alpha_s, b_1, b_2]$. All of these parameters need to be marginalised over in order to constrain the underlying cosmological parameters.
\begin{figure*}
  \centering
  \includegraphics[width=2\columnwidth]{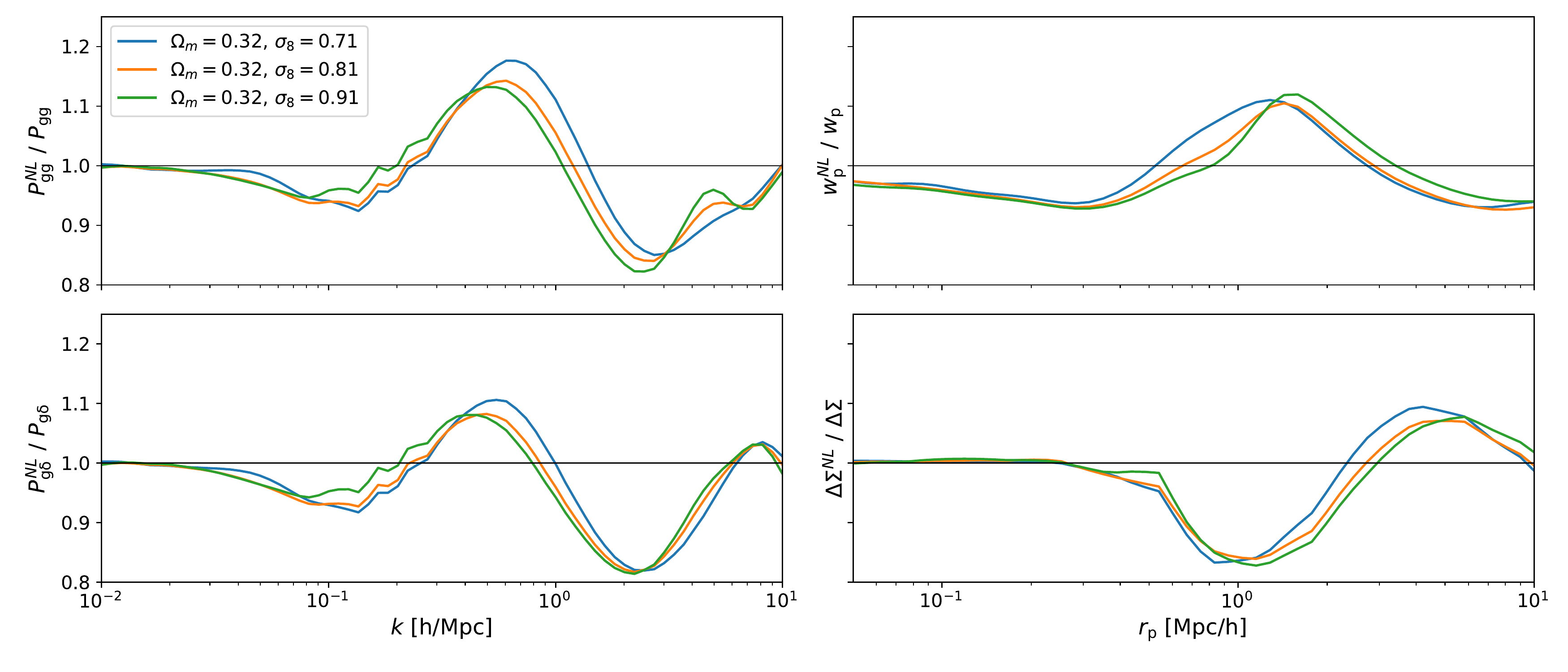}
  \caption{Fractional change in halo model predictions for lensing and clustering observables when including the beyond-linear halo bias correction $\beta^{\mathrm{NL}}$.  We present the fractional change in the galaxy-galaxy power spectrum $P_{\rm{gg}}(k)$ (upper left), projected galaxy-galaxy correlation function $w_{\rm{p}}(r_{\rm{p}})$ (upper right), galaxy-matter power spectrum $P_{\rm{g\delta}}(k)$ (lower left) and excess surface density profile $\Delta\Sigma(r_{\rm{p}})$ (lower right) at a redshift of zero for a KiDS-like survey. ${\rm NL}$ indicates that the mock data is drawn from a halo model that includes a $\Bnl$ correction to account for non-linear halo bias.  The three curves demonstrate the sensitivity of the effect to changes in $\sigma_8$ (see inset). } \label{fig:fixed_omega_m}
\end{figure*}

\begin{figure*}
  \centering
  \includegraphics[width=2\columnwidth]{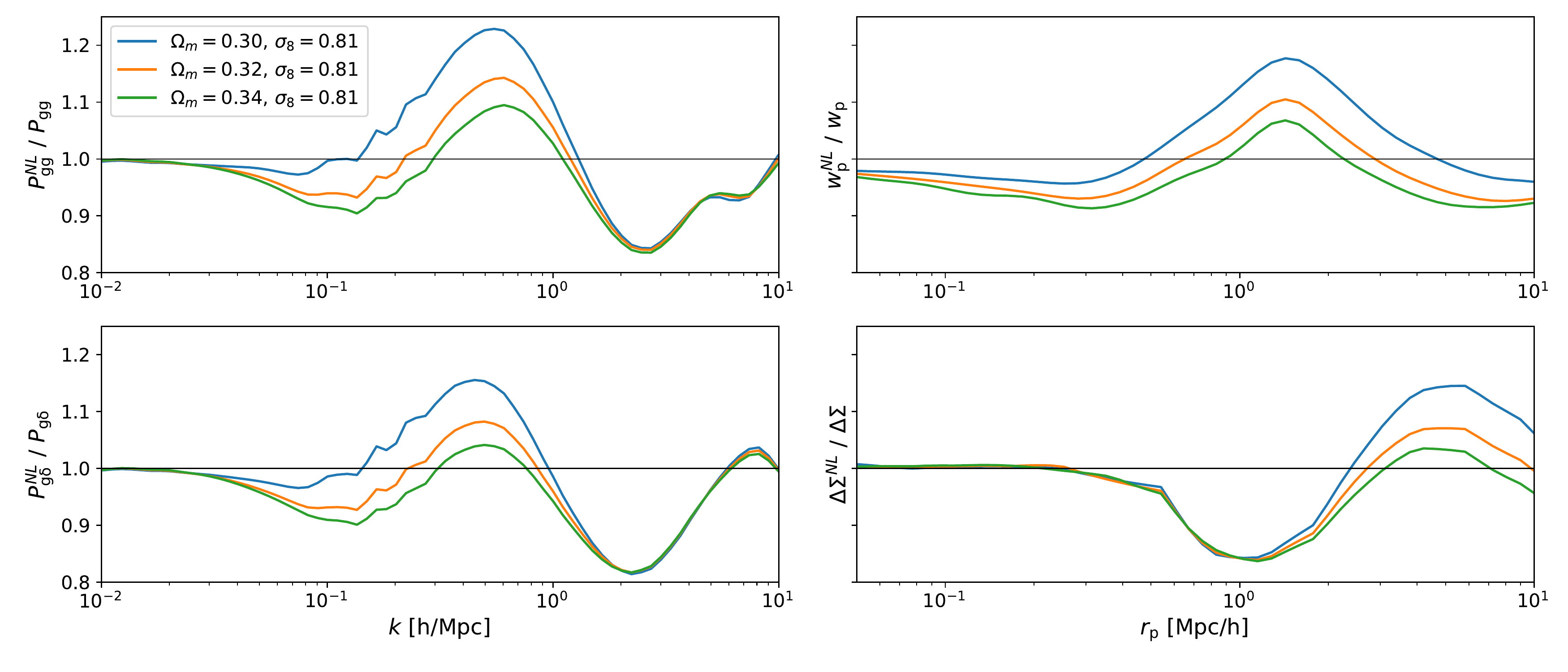}
  \caption{Fractional change in halo model predictions for lensing and clustering observables when including the beyond-linear halo bias correction $\beta^{\mathrm{NL}}$, as in Figure~\ref{fig:fixed_omega_m}.  The three curves demonstrate the sensitivity of the effect to changes in $\Omega_{\mathrm{m}}$ (see inset).}
  \label{fig:fixed_sigma8}
\end{figure*}

In Figures \ref{fig:fixed_omega_m} and \ref{fig:fixed_sigma8} we quantify the impact of including beyond-linear bias in halo model estimates of KiDS-like clustering ($P_{\rm{gg}}$, $w_{\rm{p}}$) and lensing ($P_{\rm{g\delta}}$, $\Delta\Sigma$) observables, for a range of different values for the matter density parameter, $\Omega_{\mathrm{m}}$, and the linear theory standard deviation of matter density fluctuations in a sphere of radius 8 $\mathrm{h}^{-1}$ Mpc, $\sigma_8$. In all cases we find the non-linear halo bias affects the predictions over a wide range of scales at the level of up to $\sim20$\%\footnote{We note that the curves in Figures \ref{fig:fixed_omega_m} and \ref{fig:fixed_sigma8} are not perfectly smooth.  This results from imperfections in the $\beta_{\rm NL}$ interpolation process, described in Appendix A.   Future work will optimise this interpolation procedure, but we do not anticipate these low-amplitude features to impact significantly on the findings of our analysis.}. The impact of non-linear halo bias on the two power spectra $P_{\rm{gg}}$ and $P_{\rm{g\delta}}$ (left panels) is similar, with the ratio tending to 1 on large scales as expected. These changes translate differently to the observables $w_{\rm{p}}$ and $\Delta\Sigma$ (right panels) due to projections effects. Critically, the scales impacted are those where the signal-to-noise is typically maximised in observations,  implying that there is no opportunity to mitigate the impact of beyond-linear halo bias with a halo model analysis that utilises conservative scale cuts. Focussing on the power spectra (left panels), we find the non-linear halo bias serves to increase power between $0.1 < k < 1 {\rm h \, Mpc}^{-1}$, the transition region between the one and two-halo regimes.  The addition of $\Bnl$ to our analysis therefore corrects a well documented issue with the standard halo model under-predicting the clustering in this region \citep{Tinker:2004gf,Fedeli:2014gja,Mead:2015yca}.    

Comparing Figures \ref{fig:fixed_omega_m} and \ref{fig:fixed_sigma8} we conclude that the non-linear halo bias correction is most sensitive to changes in $\Omega_{\mathrm{m}}$, with changes in the value of $\sigma_8$ making less impact.  For example, at a lower value of $\Omega_{\mathrm{m}}=0.3$, including $\beta^{\mathrm{NL}}$ in the halo model changes $P_{\rm{gg}}$ by approximately 20\%, whereas at a higher value of $\Omega_{\mathrm{m}}=0.34$, it is reduced to an approximately 10\% effect. This sensitivity has practical implications for how the $\beta^{\mathrm{NL}}$ correction can be included in cosmological analyses. Ideally $\beta^{\mathrm{NL}}$ would be emulated for each point in parameter space, but \textsc{DarkEmulator} currently has a relatively narrow cosmological range. We discuss possible solutions to this in Appendix \ref{sec:rescaling}. 

\section{Cosmological Parameters} \label{sec: cosmological parameters}

\begin{figure}
  \centering
  \includegraphics[width=1\columnwidth]{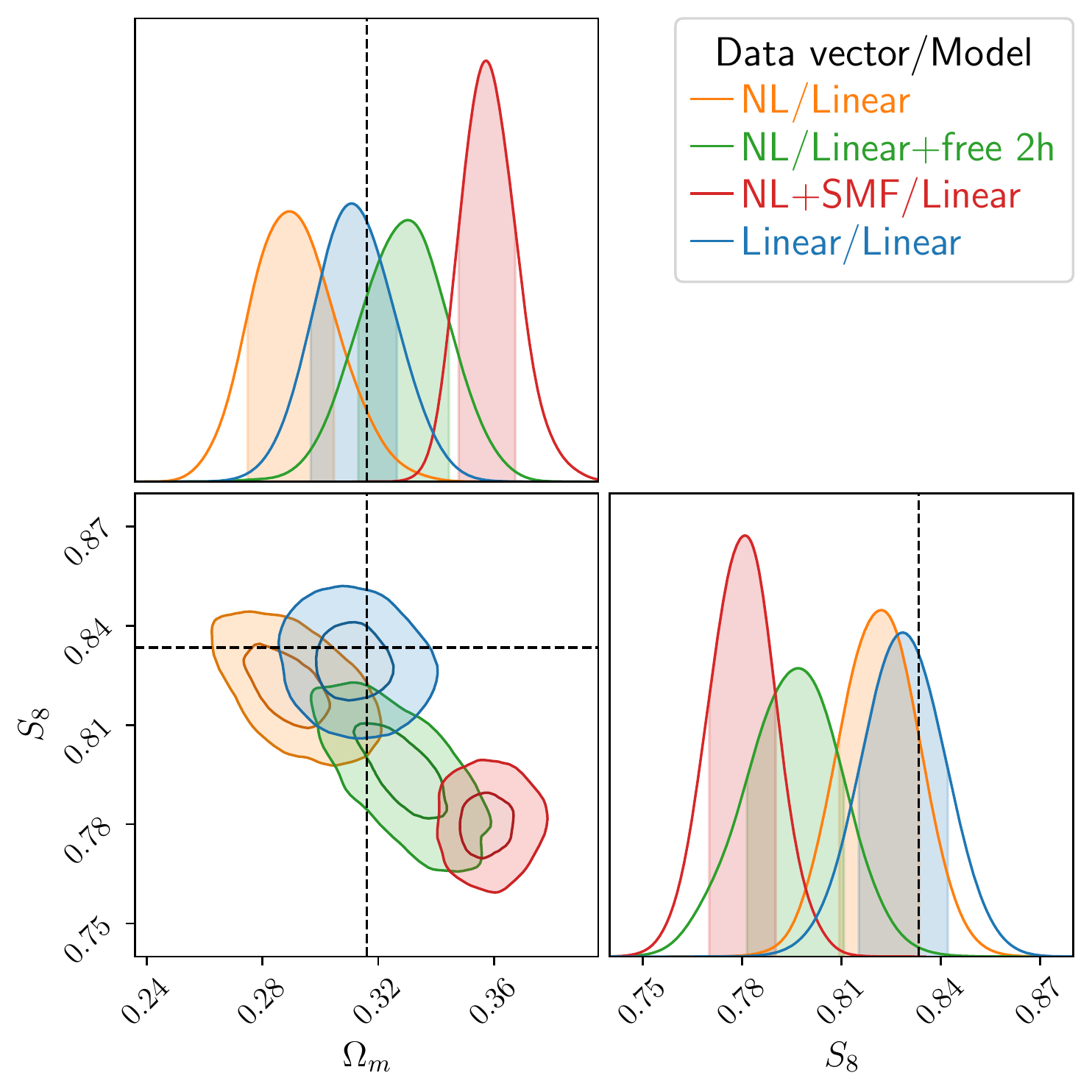}
  \caption{Forecast marginal constraints on the structure growth parameter, $S_8$, and the matter density parameter, $\Omega_{\rm m}$, for a standard halo model analysis of lensing-clustering data from a KiDS-like survey. The mock data vector is drawn from a halo model that includes a $\Bnl$ correction to account for non-linear halo bias and adopts a \citet{Planck2020} cosmology (shown dashed).   Ignoring non-linear halo bias in the halo model analysis results in an offset in the recovered cosmological constraints (orange).  This offset is not mitigated through the addition of a free multiplicative nuisance parameter (green) or through the addition of complementary stellar mass function data (SMF, red). The input cosmology is recovered when using a mock datavector drawn from a halo model which does not include a $\Bnl$ correction (blue).}
  \label{fig:S8_combined}
\end{figure}

We assess the impact of neglecting non-linear halo bias in a halo model joint lensing-clustering cosmological analysis of a KiDS-1000-like survey. Figure \ref{fig:S8_combined} presents marginalised constraints on $S_8 = \sigma_8 \sqrt{\Omega_m/0.3}$ and $\Omega_m$ for a mock joint data vector of $[w^{\rm NL}_{\rm{p},i}(r_{\rm{p}},z),\Delta\Sigma^{\rm NL}_i(r_{\rm{p}},z)]$, as shown in Figures~\ref{fig:fixed_omega_m} and \ref{fig:fixed_sigma8}. The scales included in the data vector are those shown in Figures~\ref{fig:fixed_omega_m} and \ref{fig:fixed_sigma8}, $10^{-1.3}<r_{\rm{p}} \ / h^{-1}\, \mathrm{Mpc}<10$. Here ${\rm NL}$ indicates that the mock data is drawn from a halo model that includes a $\Bnl$ correction to account for non-linear halo bias. We assume $i=1,2,3$ stellar mass bins, (10.3. 10.6], (10.6, 10.9] and (10.9, 12] with units of $\log(M_{\star} / h^{-2}\,M_{\odot})$, with a median redshift of 0.18. The mock analytical joint-covariance matrix is derived following \cite{Dvornik:2018frx,Joachimi_2021,Dvornik2022}. Our model includes 15 free parameters, 5 cosmological parameters and 10 halo model parameters (see Table \ref{tab:fiducial values}). The input fiducial cosmology is given by \citet{Planck2020} TT,TE,EE+lowE+lensing and the fiducial halo model parameters are given by \cite{vanUitert2016} (see Section \ref{sec: observables}). In this forecast we do not include modelling for intrinsic galaxy alignments or magnification, referring the reader to \citealt{Dvornik2022} where these additional terms are accounted for in the analysis. We use the Markov Chain Monte Carlo (MCMC) sampler {\sc emcee}  to explore the parameter space, and our convergence criteria is a number of samples at least 100 times the autocorrelation time \citep{emcee2013PASP..125..306F}.
\begin{table}
  \centering
  \caption{Fiducial sampling parameters and their priors. $\Omega_{\mathrm{m}}$ is the matter density parameter, $\sigma_8$ the linear theory standard deviation of matter density fluctuations in a sphere of radius 8 $\mathrm{h}^{-1}$ Mpc, $h_0$ the hubble parameter, $\Omega_{\rm{b}}$ the baryon density parameter and $n_{\rm{s}}$ the scalar spectral index. $f_{\mathrm{h}}$ and $f_{\mathrm{s}}$ normalise the concentration-mass relation for dark matter and satellite galaxies (equation \ref{eq:f_h,s}); $M_1$ is a characteristic mass scale and  $M_0$ is a normalisation; $\sigma_c$ is the scatter between stellar mass and halo mass; $\alpha_s$ governs the power law behaviour of satellite galaxies; $\gamma_1$ and $\gamma_2$ are powers in the expression for the stellar mass of centrals; and $b_1$ and $b_2$ enter the expression for the satellite stellar mass function (equations \ref{phi_c}-\ref{eq:CMF7}).}
  \begin{tabular}{lll}
  \textbf{Parameter} & \textbf{Fiducial Value} & \textbf{Prior}  \\
  \hline
  \hline
            & \textbf{Cosmology} &  \\
       $\Omega_{\rm{m}}$     &  $0.3158$ & [0.1, 0.45] \\
       $\sigma_8$ & $0.812$ & [0.6, 1.0] \\
            $h_0$     &  $0.6732$ & [0.64, 0.82] \\
            $\Omega_{\rm{b}}$     & $0.0494$ & [0.01, 0.06] \\
            $n_{\rm{s}}$     & $0.9661$ & [0.84, 1.1] \\
  \hline
            & \textbf{CLF}&  \\
            $f_{\mathrm{h}}$     & $1.0$ & [0.0, 1.2] \\
            $\log(M_0)$     & $10.58$ & [9.0, 13.0] \\
            $\log(M_1)$     & $10.97$ & [9.0, 14.0] \\
            $\gamma_1$     & $7.5$ & [5.5, 9.5] \\
            $\gamma_2$     & $0.25$ & [0.001, 1.0] \\
            $\sigma_c$     & $0.2$ & [0.1, 1.0] \\           
            $f_s$     & $1.0$ & [0.0, 1.2] \\
            $\alpha_s$     & $-0.83$ & [-1.1, -0.6] \\
            $b_1$     & $0.18$ & [-0.2, 0.3] \\
            $b_2$     & $0.83$ & [0.6, 0.9] \\
  \end{tabular}
  \label{tab:fiducial values}
  \end{table} 

Figure \ref{fig:S8_combined} compares four different cosmological analyses. In all cases, the analysis pipeline assumes linear halo bias with $I_{\mathrm{xy}}^{\mathrm{NL}}(k,z)=0$ in equation~\ref{eq:2 halo term}, and the data vector either includes non-linear halo bias (NL) or in the case of the blue contours is matched to the analysis pipeline. For the blue contours we expect to recover the input cosmology and any differences are due to projection effects when marginalising over many parameters (see for example \citealt{Joachimi_2021}). We find that the marginal constraints on $S_8=0.829$ and $\Omega_{\rm m}=0.311$, which are offset with respect to the input by $0.4\sigma$ and $0.3\sigma$, respectively. We take these small projection effects into account when estimating offsets in parameters for the remaining cases, and use the marginal distributions to compute the offsets. We have verified that the best fit values for all cases are close to the maximum of the marginal distributions. Adopting a standard halo model analysis results in a $1.4\sigma$ offset in the recovered value of $\Omega_m$ (orange contour). Introducing an additional free nuisance parameter to the standard halo model analysis $a$, allowing for freedom in the amplitude of the central and satellite two-halo power spectra (e.g. $P_{\rm{cs}}^{\rm{2h}}\rightarrow a^2 P_{\rm{cs}}^{\rm{2h}}, P_{\rm{c\delta}}^{\rm{2h}}\rightarrow aP_{\rm{c\delta}}^{\rm{2h}}$), resolves some of the offset in $\Omega_{\rm m}$.  As this additional power parameter is degenerate with $\sigma_8$, however, the inclusion of this multiplicative nuisance term $a$ to account for an additive astrophysical systematic results in a $2.3\sigma$ offset in the recovered value of $S_8$ (green contour).

\begin{figure*}
  \centering
  \includegraphics[width=2\columnwidth]{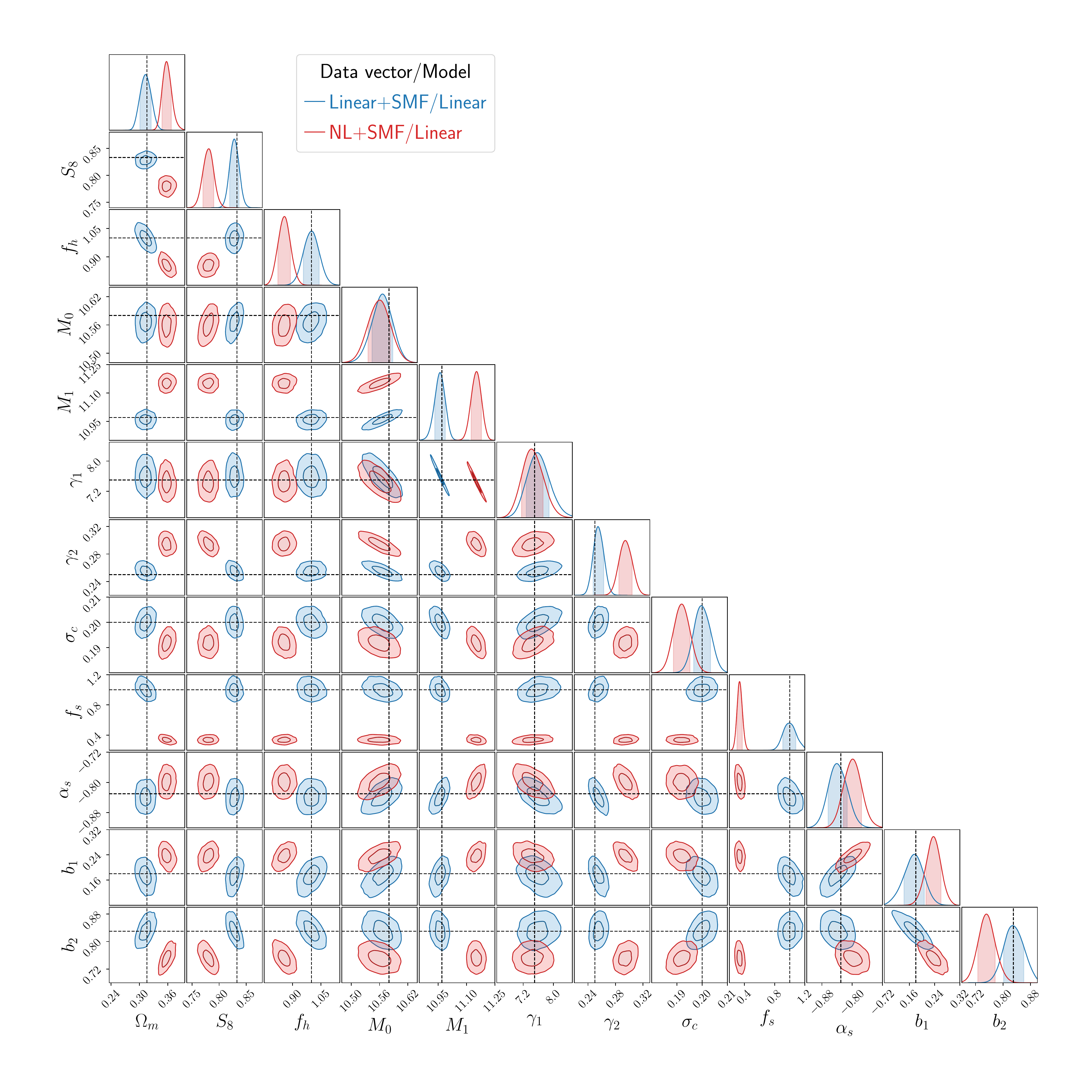}
  \caption{Forecast marginal constraints on the structure growth parameter, $S_8$, the matter density parameter, $\Omega_{\rm m}$, and the conditional stellar mass function (CSMF) parameters for a standard halo model analysis of lensing-clustering and stellar mass function (SMF) data from a KiDS-like survey. The mock data vector is drawn from a halo model that includes a $\Bnl$ correction to account for non-linear halo bias (red) in comparison to a reference case with linear halo bias (blue). Ignoring non-linear halo bias in the halo model analysis results in particularly large offsets in both the CSMF and cosmological parameters when including the SMF, as the SMF breaks degeneracies between the cosmological and CSMF parameter when combined with lensing and clustering. $f_{\mathrm{h}}$ and $f_{\mathrm{s}}$ normalise the concentration-mass relation for dark matter and satellite galaxies (equation \ref{eq:f_h,s}); $M_1$ is a characteristic mass scale and $M_0$ is a normalisation; $\sigma_c$ is the scatter between stellar mass and halo mass; $\alpha_s$ governs the power law behaviour of satellite galaxies; $\gamma_1$ and $\gamma_2$ are powers in the expression for the stellar mass of centrals; and $b_1$ and $b_2$ enter the expression for the satellite stellar mass function (equations \ref{phi_c}-\ref{eq:CMF7}).}
  \label{fig:SMF}
\end{figure*}

Observations of the stellar mass function (SMF) are known to enhance constraints on the halo model parameters \citep{vanUitert2016}.  We find that the inclusion of a mock SMF into our data vector results in the largest offset in the recovered cosmological parameters, with a $5.2\sigma$ $S_8$ offset and a $4.9\sigma$ $\Omega_{\rm m}$ offset (red contour). Here including the SMF breaks degeneracies between the cosmological parameters and the CSMF parameters, which determine the central and satellite profiles $\mathcal{H}_{\rm x}$ \citep{More_2013}. This tightens the parameter constraints and results in greater offsets. Figure \ref{fig:SMF} shows the most significant offset is in the value of $f_s$ the normalisation of the concentration-mass relation for satellite galaxies (equation \ref{eq:f_h,s}). This follows as $f_s$ is not constrained by the SMF but by the lensing and clustering, which are missing the non-linear halo bias. In contrast $M_0$ and $\gamma_1$ are less affected as they are predominantly determined by the high stellar mass region, which is largely constrained by the SMF. Including the SMF can therefore be very useful in a joint lensing-clustering halo model analysis, but only if the halo model is fully representative of the underlying observables.

Looking at the reduced chi-squared for the three different cosmological analyses in Figure \ref{fig:S8_combined}, we find that all provide a good fit to the data. It is therefore not feasible to assess the model using goodness of fit, but the offsets in the cosmological parameters clearly show that non-linear halo bias can no longer be neglected in galaxy-galaxy lensing and galaxy clustering halo model analyses.

\subsection{Comparison to \textsc{DarkEmulator}}
\label{sec:DEcomp}

\citet{miyatake/etal:2021} present cosmological parameter constraints from a joint lensing-clustering analysis of the Hyper Suprime-Cam Survey \citep[HSC,][]{HSC_2017} and the Baryon Oscillation Spectroscopic Survey \citep[BOSS,][]{BOSS/etal:2013}.    Utilising the \textsc{DarkEmulator} they extract direct measurements of the halo-matter cross-power spectrum $\hat{P}_{\rm hm}(k)$ and the halo-halo power spectrum $\hat{P}_{\rm hh}(k)$ from the Dark Quest simulations \citep{Miyatake:2020uhg}.  In doing so they bypass the traditional halo-model route of constructing these quantities using simulation-calibrated fitting functions of the halo mass function, $n(M,z)$, the halo bias, $b(M,z)$, and the halo density profile  \citep{2010tinker,NFW_1997,Duffy_2008}.   This approach therefore automatically accounts for the non-linear halo bias and halo exclusion, that we have encapsulated with the \textsc{DarkEmulator} estimates of $\Bnl$.  As Dark Quest is a dark-matter only simulation, \citet{miyatake/etal:2021} then use a halo occupation distribution (HOD) to map the galaxy-halo connection and predict $w_p(r)$ and $\Delta\Sigma(r)$ observables to compare with HSC observations and set tight constraints on $S_8$ and $\Omega_{\rm m}$.

\begin{figure}
  \centering
  \includegraphics[width=1.0\columnwidth]{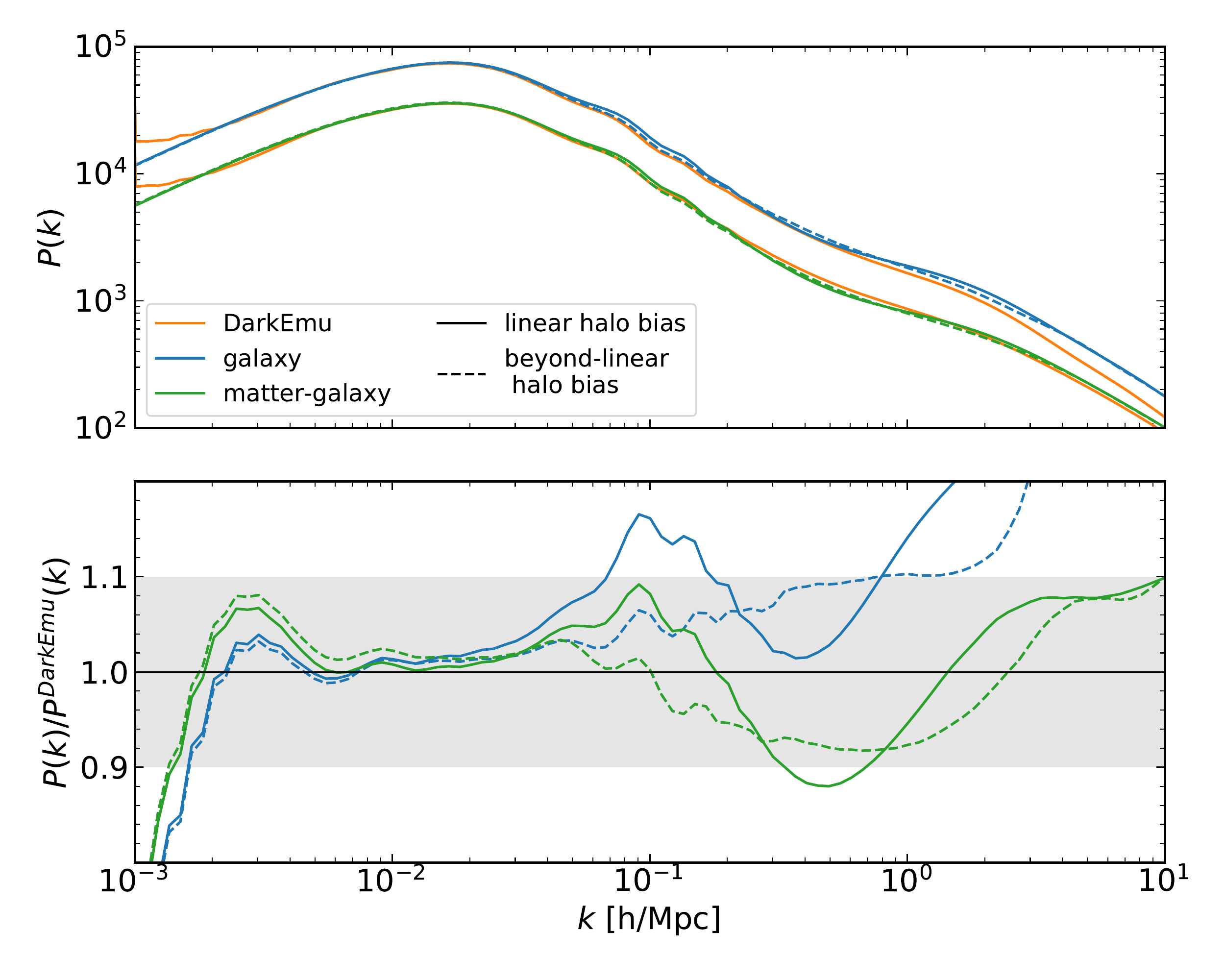} 
  \caption{
   Comparision between the galaxy-galaxy and galaxy-matter power spectra computed directly from \textsc{DarkEmulator}+HOD~\citep{DarkEmulatorCode,Miyatake:2020uhg} and from our halo model with and without the beyond-linear halo bias correction $\beta^{\mathrm{NL}}$. We find broad consistency, at the level of $\sim 10\%$ (grey bar).
}
  \label{fig:daremu_comp}
\end{figure}

Figure~\ref{fig:daremu_comp} compares the emulated galaxy-galaxy and galaxy-matter power spectra from \citet{Miyatake:2020uhg} to the two halo model approaches presented in Section~\ref{sec: observables}.  Here we match the HOD galaxy-halo prescription, simulating a BOSS-like sample of luminous red galaxies.   Any differences in the models therefore arise from the different approaches taken to determine the underlying halo-matter connection.  We find consistency with broad agreement within $\sim10\%$ accuracy (grey band).   The inclusion of the $\Bnl$ non-linear halo bias correction (dashed) is shown to improve the agreement particularly around the transition region, $ k \sim 0.1 {\rm h \, Mpc}^{-1}$.   On very small and large scales, we note that simulation resolution and sampling effects come into play with Dark Quest (see the discussion in Appendix~\ref{app:DarkEmu}).

\begin{figure}
  \centering
  \includegraphics[width=1.0\columnwidth]{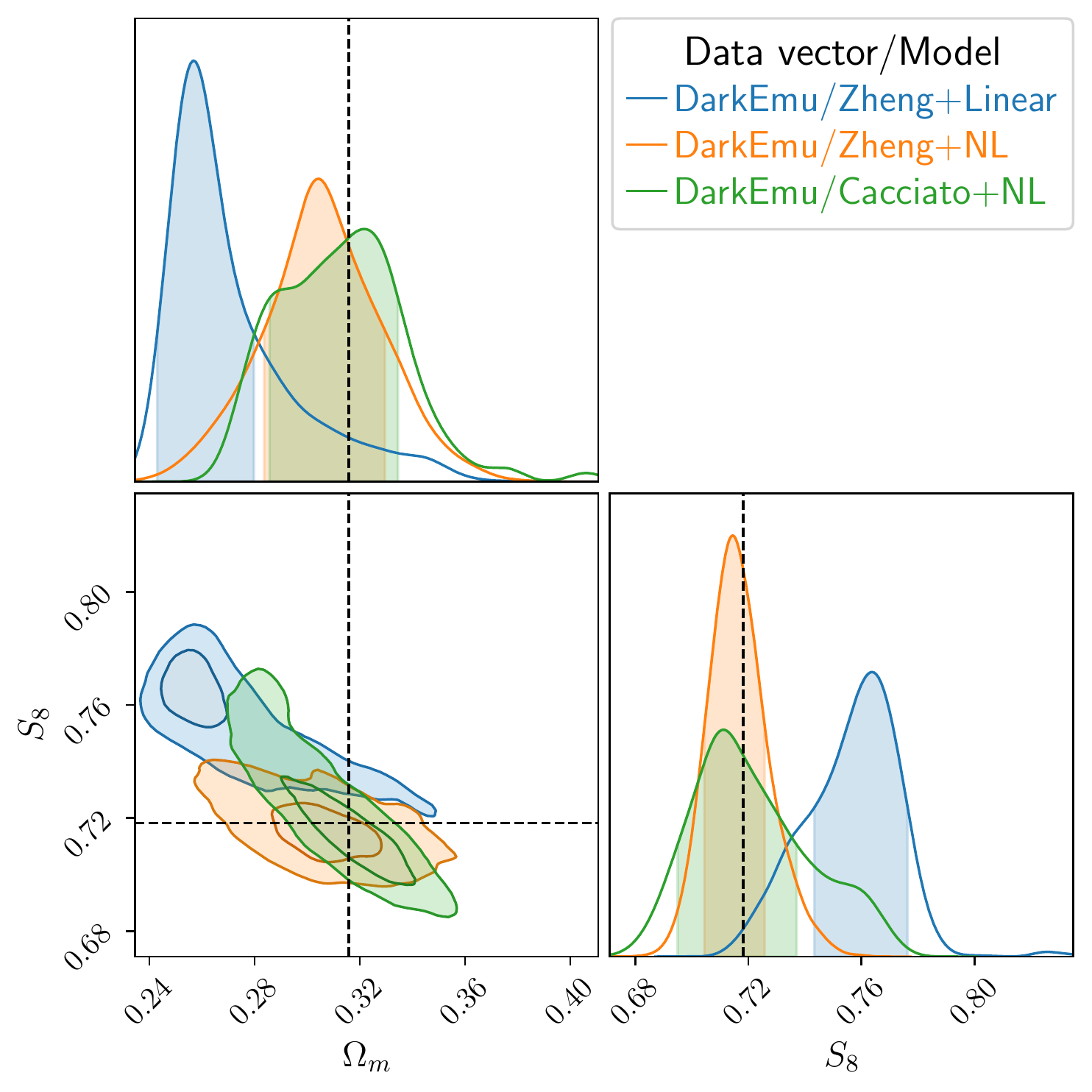} 
  \caption{Forecast marginal constraints on the structure growth parameter, $S_8$, and the matter density parameter, $\Omega_{\rm m}$, for a standard halo model analysis with linear halo bias (blue) and a halo model analysis that includes a $\Bnl$ correction to account for non-linear halo bias (orange, green). The mock lensing-clustering data is drawn from \textsc{DarkEmulator}+HOD~\citep{DarkEmulatorCode,Miyatake:2020uhg}. The $\sim 10\%$ differences shown in Figure~\ref{fig:daremu_comp} do not translate into offsets in the cosmological parameters when including $\Bnl$. The input cosmology (shown dashed) is recovered when matching the \citet{Zheng:2004id} HOD model utilised by \textsc{DarkEmulator} (orange) and when using the \citet{Cacciato2013MNRAS.430..767C} HOD model adopted in the rest of this analysis (green).}
  \label{fig:daremu_datavector}
\end{figure}

In Figure~\ref{fig:daremu_datavector} we perform a cosmological analysis of the {\sc DarkEmulator} mock $w_p(r)$ and $\Delta\Sigma(r)$ observables, assuming KiDS-like errors.  Similar to the findings in Figure~\ref{fig:S8_combined}, we recover a significant offset in the recovered parameters when assuming linear halo bias in our halo model with a $2.4\sigma$ offset in $S_8$ and a $2.1\sigma$ offset in $\Om$\footnote{We compute the offset for the linear halo bias case (blue contour) with respect to the non-linear bias case (orange contour), using the same approach as for Figure~\ref{fig:S8_combined}.}. When including the non-linear bias model $\Bnl$, however, we find that the halo model is flexible enough to recover the input cosmology, despite the $\sim10\%$ differences in Figure~\ref{fig:daremu_comp}. This is true when matching the \citet{Zheng:2004id} HOD galaxy-halo prescription utilised with \textsc{DarkEmulator}, which does not include stellar masses, and when using the \citet{Cacciato2013MNRAS.430..767C} HOD prescription utilised in the rest of this work (Figures \ref{fig:fixed_omega_m}-\ref{fig:SMF}).

\citet{Miyatake:2020uhg} conduct a similar experiment to Figure \ref{fig:daremu_datavector}, determining the offset in the recovered cosmological parameter constraints when analysing a mock data vector from the \textsc{DarkEmulator}+HOD with a fully analytic halo model. They find offsets in $S_8$ and $\Om$, with similar directions to Figure \ref{fig:daremu_datavector}, but with reduced magnitudes due to their wider uncertainties. Figures \ref{fig:S8_combined} and \ref{fig:daremu_datavector}, however, show different offsets in the recovered cosmologies. This is further confirmation that the halo model configuration choice can have a significant impact on the final parameter biases in the $S_8-\Omega_{\rm m}$ space, as demonstrated by the other contours in Figure \ref{fig:S8_combined}. We therefore conclude from these studies that cosmological parameter constraints are sensitive to missing ingredients and how they are accounted for within the halo model. 

\section{Conclusions} \label{sec: conclusions}
In this paper we review the accuracy of cosmological parameter constraints from a joint halo model analysis of galaxy-galaxy lensing and clustering in a KiDS-like survey.   We find that significant offsets, up to $\sim 5\sigma$ level, are introduced in the marginal constraints on $S_8$ and $\Om$, when taking the standard halo model approach of neglecting the non-linear bias of halos.   We adopt the beyond-linear halo bias correction $\beta^{\mathrm{NL}}$, proposed in \cite{Mead:2020qdk}, which we re-calibrate using the Dark Quest simulations (see Appendix \ref{app:DarkEmu}).  We find that the amplitude of the projected galaxy-galaxy correlation function $w_{\rm{p}}(r_{\rm{p}})$, and the excess surface density profile $\Delta\Sigma(r_{\rm{p}})$, are impacted at the level of up to $\sim 20\%$.  Importantly, neglecting the non-linear halo bias impacts a very wide range of scales such that there is little opportunity to mitigate this approximation using scale-cuts.   We therefore conclude that any future halo model large-scale structure study must include non-linear halo bias modelling in their analysis.

In a recent joint HSC-BOSS galaxy-galaxy lensing and clustering analysis, \citet{miyatake/etal:2021} employ the N-body simulations from the {\sc DarkEmulator} to model their observables \citep{Nishimichi2019, Miyatake:2020uhg}.  This approach naturally incorporates non-linear halo bias, as the halos are extracted directly from the simulations.   We demonstrate that a halo model analysis of KiDS-like {\sc DarkEmulator} galaxy-galaxy lensing and clustering observables can accurately recover the input cosmology, at the level of $0.3\sigma$, provided the \citet{Mead:2020qdk} $\beta^{\mathrm{NL}}$ correction is included in the analysis.

One benefit of employing the Dark Quest calibrated $\beta^{\mathrm{NL}}$ correction, in contrast to a direct emulation of observables with {\sc DarkEmulator}, is the retention of the halo model flexibility.  Importantly the halo model allows for the marginalisation over nuisance parameters that can account for uncertainty on the impact of baryon feedback on the simulated dark matter distribution \citep{Debackere:2021ado}.  This approach also facilitates straightforward extensions to simultaneously model multiple large-scale structure probes and constrain baryon feedback models \citep[see for example][]{HMxCode,Acuto_2021,Troester_2021}.  In addition a halo model approach permits the study of an essentially unlimited range of exotic cosmological models \citep{Cataneo_2019, Bose_2020}.

One caveat to this work is the existence of `assembly bias', a term which refers to the assumption in halo modelling that the clustering of halos depends only on their mass and not their assembly history \citep{Gao:2005ca,Wechsler:2005gb,Dalal:2008zd}. We do not account for our uncertainty over the significance of assembly bias in our analysis, but refer to \cite{Miyatake:2020uhg} where they find that even in a maximum assembly bias scenario, the impact of assembly bias can largely be mitigated by scale cuts.

\section*{Acknowledgements}

We thank and acknowledge support from the European Research Council under grant agreement No.~770935 (CM, AD, HH, RR) and No.~647112 (AM, CH, and MA).  We also acknowledge support from the Max Planck Society and the Alexander von Humboldt Foundation in the framework of the Max Planck-Humboldt Research Award endowed by the Federal Ministry of Education and Research (AM, CH). 
TN and HM were supported in part by World Premier International Research Center Initiative (WPI Initiative), MEXT, Japan, Japan Science and Technology Agency (JST) AIP Acceleration Research Grant Number JP20317829, MEXT/JSPS KAKENHI Grant Number JP19H00677, and MEXT/JSPS Core-to-Core Program Grant Number JPJSCCA20200002. TN was supported in part by MEXT/JSPS KAKENHI Grant Numbers JP20H05861 and JP21H01081. HM was supported in part by MEXT/JSPS KAKENHI Grant Numbers JP20H01932 and JP21H05456.

\section*{Data Availability}

The \textsc{DarkEmulator} is publicly available here: \url{https://github.com/DarkQuestCosmology/dark_emulator_public}



\bibliographystyle{mnras}
\bibliography{beyond_linear_halo_bias} 

\begin{thebibliography}{}
\makeatletter
\relax
\def\mn@urlcharsother{\let\do\@makeother \do\$\do\&\do\#\do\^\do\_\do\%\do\~}
\def\mn@doi{\begingroup\mn@urlcharsother \@ifnextchar [ {\mn@doi@}
  {\mn@doi@[]}}
\def\mn@doi@[#1]#2{\def\@tempa{#1}\ifx\@tempa\@empty \href
  {http://dx.doi.org/#2} {doi:#2}\else \href {http://dx.doi.org/#2} {#1}\fi
  \endgroup}
\def\mn@eprint#1#2{\mn@eprint@#1:#2::\@nil}
\def\mn@eprint@arXiv#1{\href {http://arxiv.org/abs/#1} {{\tt arXiv:#1}}}
\def\mn@eprint@dblp#1{\href {http://dblp.uni-trier.de/rec/bibtex/#1.xml}
  {dblp:#1}}
\def\mn@eprint@#1:#2:#3:#4\@nil{\def\@tempa {#1}\def\@tempb {#2}\def\@tempc
  {#3}\ifx \@tempc \@empty \let \@tempc \@tempb \let \@tempb \@tempa \fi \ifx
  \@tempb \@empty \def\@tempb {arXiv}\fi \@ifundefined
  {mn@eprint@\@tempb}{\@tempb:\@tempc}{\expandafter \expandafter \csname
  mn@eprint@\@tempb\endcsname \expandafter{\@tempc}}}

\bibitem[\protect\citeauthoryear{{Acuto}, {McCarthy}, {Kwan}, {Salcido},
  {Stafford}  \& {Font}}{{Acuto} et~al.}{2021}]{Acuto_2021}
{Acuto} A.,  {McCarthy} I.~G.,  {Kwan} J.,  {Salcido} J.,  {Stafford} S.~G.,
  {Font} A.~S.,  2021, \mn@doi [\mnras] {10.1093/mnras/stab2834}, \href
  {https://ui.adsabs.harvard.edu/abs/2021MNRAS.tmp.2595A} {}

\bibitem[\protect\citeauthoryear{Aihara et~al.,}{Aihara
  et~al.}{2017}]{HSC_2017}
Aihara H.,  et~al., 2017, \mn@doi [Publications of the Astronomical Society of
  Japan] {10.1093/pasj/psx066}, 70

\bibitem[\protect\citeauthoryear{Alam et~al.}{Alam et~al.}{2017}]{BOSS:2016wmc}
Alam S.,  et~al., 2017, \mn@doi [Mon. Not. Roy. Astron. Soc.]
  {10.1093/mnras/stx721}, 470, 2617

\bibitem[\protect\citeauthoryear{Alam et~al.}{Alam
  et~al.}{2021}]{eBOSS:2020yzd}
Alam S.,  et~al., 2021, \mn@doi [Phys. Rev. D] {10.1103/PhysRevD.103.083533},
  103, 083533

\bibitem[\protect\citeauthoryear{{Angulo} \& {Hilbert}}{{Angulo} \&
  {Hilbert}}{2015}]{Angulo2015}
{Angulo} R.~E.,  {Hilbert} S.,  2015, \mn@doi [\mnras] {10.1093/mnras/stv050},
  \href {http://adsabs.harvard.edu/abs/2015MNRAS.448..364A} {448, 364}

\bibitem[\protect\citeauthoryear{{Angulo} \& {White}}{{Angulo} \&
  {White}}{2010}]{Angulo2010}
{Angulo} R.~E.,  {White} S.~D.~M.,  2010, \mn@doi [\mnras]
  {10.1111/j.1365-2966.2010.16459.x}, \href
  {http://adsabs.harvard.edu/abs/2010MNRAS.405..143A} {405, 143}

\bibitem[\protect\citeauthoryear{{Aric{\`o}}, {Angulo},
  {Hern{\'a}ndez-Monteagudo}, {Contreras}, {Zennaro}, {Pellejero-Iba{\~n}ez}
  \& {Rosas-Guevara}}{{Aric{\`o}} et~al.}{2020}]{Giovanni2020}
{Aric{\`o}} G.,  {Angulo} R.~E.,  {Hern{\'a}ndez-Monteagudo} C.,  {Contreras}
  S.,  {Zennaro} M.,  {Pellejero-Iba{\~n}ez} M.,   {Rosas-Guevara} Y.,  2020,
  \mn@doi [\mnras] {10.1093/mnras/staa1478}, \href
  {https://ui.adsabs.harvard.edu/abs/2020MNRAS.495.4800A} {495, 4800}

\bibitem[\protect\citeauthoryear{Baldauf, Seljak, Smith, Hamaus  \&
  Desjacques}{Baldauf et~al.}{2013}]{Baldauf_2013}
Baldauf T.,  Seljak U.,  Smith R.~E.,  Hamaus N.,   Desjacques V.,  2013,
  \mn@doi [Physical Review D] {10.1103/physrevd.88.083507}, 88

\bibitem[\protect\citeauthoryear{Behroozi, Wechsler  \& Wu}{Behroozi
  et~al.}{2012}]{Rockstar}
Behroozi P.~S.,  Wechsler R.~H.,   Wu H.-Y.,  2012, \mn@doi [The Astrophysical
  Journal] {10.1088/0004-637x/762/2/109}, 762, 109

\bibitem[\protect\citeauthoryear{Bilicki et~al.}{Bilicki
  et~al.}{2021}]{Bilicki:2021hgn}
Bilicki M.,  et~al., 2021, \mn@doi [Astron. Astrophys.]
  {10.1051/0004-6361/202140352}, 653, A82

\bibitem[\protect\citeauthoryear{{Bond}, {Cole}, {Efstathiou}  \&
  {Kaiser}}{{Bond} et~al.}{1991}]{Bond1991}
{Bond} J.~R.,  {Cole} S.,  {Efstathiou} G.,   {Kaiser} N.,  1991, \mn@doi
  [\apj] {10.1086/170520}, \href
  {http://adsabs.harvard.edu/abs/1991ApJ...379..440B} {379, 440}

\bibitem[\protect\citeauthoryear{{Bose}, {Cataneo}, {Tr{\"o}ster}, {Xia},
  {Heymans}  \& {Lombriser}}{{Bose} et~al.}{2020}]{Bose_2020}
{Bose} B.,  {Cataneo} M.,  {Tr{\"o}ster} T.,  {Xia} Q.,  {Heymans} C.,
  {Lombriser} L.,  2020, \mn@doi [\mnras] {10.1093/mnras/staa2696}, \href
  {https://ui.adsabs.harvard.edu/abs/2020MNRAS.498.4650B} {498, 4650}

\bibitem[\protect\citeauthoryear{Bullock, Kolatt, Sigad, Somerville, Kravtsov,
  Klypin, Primack  \& Dekel}{Bullock et~al.}{2001}]{Bullock2001}
Bullock J.~S.,  Kolatt T.~S.,  Sigad Y.,  Somerville R.~S.,  Kravtsov A.~V.,
  Klypin A.~A.,  Primack J.~R.,   Dekel A.,  2001, \mn@doi [\mnras]
  {10.1046/j.1365-8711.2001.04068.x}, 321, 559

\bibitem[\protect\citeauthoryear{Cacciato, Bosch, More, Li, Mo  \&
  Yang}{Cacciato et~al.}{2009}]{Cacciato:2008hm}
Cacciato M.,  Bosch F. C. v.~d.,  More S.,  Li R.,  Mo H.~J.,   Yang X.,  2009,
  \mn@doi [Mon. Not. Roy. Astron. Soc.] {10.1111/j.1365-2966.2008.14362.x},
  394, 929

\bibitem[\protect\citeauthoryear{{Cacciato}, {Lahav}, {van den Bosch},
  {Hoekstra}  \& {Dekel}}{{Cacciato} et~al.}{2012}]{Cacciato2012}
{Cacciato} M.,  {Lahav} O.,  {van den Bosch} F.~C.,  {Hoekstra} H.,   {Dekel}
  A.,  2012, \mn@doi [\mnras] {10.1111/j.1365-2966.2012.21762.x}, \href
  {https://ui.adsabs.harvard.edu/abs/2012MNRAS.426..566C} {426, 566}

\bibitem[\protect\citeauthoryear{{Cacciato}, {van den Bosch}, {More}, {Mo}  \&
  {Yang}}{{Cacciato} et~al.}{2013}]{Cacciato2013MNRAS.430..767C}
{Cacciato} M.,  {van den Bosch} F.~C.,  {More} S.,  {Mo} H.,   {Yang} X.,
  2013, \mnras, \href {https://ui.adsabs.harvard.edu/abs/2013MNRAS.430..767C}
  {430, 767}

\bibitem[\protect\citeauthoryear{{Cataneo}, {Lombriser}, {Heymans}, {Mead},
  {Barreira}, {Bose}  \& {Li}}{{Cataneo} et~al.}{2019}]{Cataneo_2019}
{Cataneo} M.,  {Lombriser} L.,  {Heymans} C.,  {Mead} A.~J.,  {Barreira} A.,
  {Bose} S.,   {Li} B.,  2019, \mn@doi [\mnras] {10.1093/mnras/stz1836}, \href
  {https://ui.adsabs.harvard.edu/abs/2019MNRAS.488.2121C} {488, 2121}

\bibitem[\protect\citeauthoryear{Contreras, Angulo, Zennaro, Aric\`o  \&
  Pellejero-Iba\~nez}{Contreras et~al.}{2020}]{Contreras2020}
Contreras S.,  Angulo R.~E.,  Zennaro M.,  Aric\`o G.,   Pellejero-Iba\~nez M.,
   2020, \mn@doi [Mon. Not. Roy. Astron. Soc.] {10.1093/mnras/staa3117}, 499,
  4905

\bibitem[\protect\citeauthoryear{Cooray \& Sheth}{Cooray \&
  Sheth}{2002}]{Cooray:2002dia}
Cooray A.,  Sheth R.~K.,  2002, \mn@doi [Phys. Rept.]
  {10.1016/S0370-1573(02)00276-4}, 372, 1

\bibitem[\protect\citeauthoryear{Crocce \& Scoccimarro}{Crocce \&
  Scoccimarro}{2006}]{Crocce_2006}
Crocce M.,  Scoccimarro R.,  2006, \mn@doi [Physical Review D]
  {10.1103/physrevd.73.063520}, 73

\bibitem[\protect\citeauthoryear{Dalal, White, Bond  \& Shirokov}{Dalal
  et~al.}{2008}]{Dalal:2008zd}
Dalal N.,  White M.,  Bond J.~R.,   Shirokov A.,  2008, \mn@doi [Astrophys. J.]
  {10.1086/591512}, 687, 12

\bibitem[\protect\citeauthoryear{{Dawson} et~al.,}{{Dawson}
  et~al.}{2013}]{BOSS/etal:2013}
{Dawson} K.~S.,  et~al., 2013, \mn@doi [\aj] {10.1088/0004-6256/145/1/10},
  \href {https://ui.adsabs.harvard.edu/abs/2013AJ....145...10D} {145, 10}

\bibitem[\protect\citeauthoryear{Debackere, Schaye  \& Hoekstra}{Debackere
  et~al.}{2021}]{Debackere:2021ado}
Debackere S. N.~B.,  Schaye J.,   Hoekstra H.,  2021, \mn@doi [Mon. Not. Roy.
  Astron. Soc.] {10.1093/mnras/stab1326}, 505, 593

\bibitem[\protect\citeauthoryear{Dekel \& Lahav}{Dekel \&
  Lahav}{1999}]{Dekel:1998eq}
Dekel A.,  Lahav O.,  1999, \mn@doi [Astrophys. J.] {10.1086/307428}, 520, 24

\bibitem[\protect\citeauthoryear{{Despali}, {Giocoli}, {Angulo}, {Tormen},
  {Sheth}, {Baso}  \& {Moscardini}}{{Despali} et~al.}{2016}]{Despali2016}
{Despali} G.,  {Giocoli} C.,  {Angulo} R.~E.,  {Tormen} G.,  {Sheth} R.~K.,
  {Baso} G.,   {Moscardini} L.,  2016, \mn@doi [\mnras]
  {10.1093/mnras/stv2842}, \href
  {http://adsabs.harvard.edu/abs/2016MNRAS.456.2486D} {456, 2486}

\bibitem[\protect\citeauthoryear{{Driver} et~al.,}{{Driver}
  et~al.}{2011}]{driver2011MNRAS.413..971D}
{Driver} S.~P.,  et~al., 2011, \mn@doi [\mnras]
  {10.1111/j.1365-2966.2010.18188.x}, \href
  {https://ui.adsabs.harvard.edu/abs/2011MNRAS.413..971D} {413, 971}

\bibitem[\protect\citeauthoryear{{Duffy}, {Schaye}, {Kay}  \& {Dalla
  Vecchia}}{{Duffy} et~al.}{2008}]{Duffy_2008}
{Duffy} A.~R.,  {Schaye} J.,  {Kay} S.~T.,   {Dalla Vecchia} C.,  2008, \mn@doi
  [\mnras] {10.1111/j.1745-3933.2008.00537.x}, \href
  {https://ui.adsabs.harvard.edu/abs/2008MNRAS.390L..64D} {390, L64}

\bibitem[\protect\citeauthoryear{Dvornik et~al.}{Dvornik
  et~al.}{2018}]{Dvornik:2018frx}
Dvornik A.,  et~al., 2018, \mn@doi [Mon. Not. Roy. Astron. Soc.]
  {10.1093/mnras/sty1502}, 479, 1240

\bibitem[\protect\citeauthoryear{Dvornik et~al.}{Dvornik
  et~al.}{prep}]{Dvornik2022}
Dvornik A.,  et~al., in prep.

\bibitem[\protect\citeauthoryear{Eisenstein \& Hu}{Eisenstein \&
  Hu}{1998}]{Eisenstein:1997ik}
Eisenstein D.~J.,  Hu W.,  1998, \mn@doi [Astrophys. J.] {10.1086/305424}, 496,
  605

\bibitem[\protect\citeauthoryear{Fedeli, Semboloni, Velliscig, Van~Daalen,
  Schaye  \& Hoekstra}{Fedeli et~al.}{2014}]{Fedeli:2014gja}
Fedeli C.,  Semboloni E.,  Velliscig M.,  Van~Daalen M.,  Schaye J.,   Hoekstra
  H.,  2014, \mn@doi [JCAP] {10.1088/1475-7516/2014/08/028}, 08, 028

\bibitem[\protect\citeauthoryear{{Foreman-Mackey}, {Hogg}, {Lang}  \&
  {Goodman}}{{Foreman-Mackey} et~al.}{2013}]{emcee2013PASP..125..306F}
{Foreman-Mackey} D.,  {Hogg} D.~W.,  {Lang} D.,   {Goodman} J.,  2013, \mn@doi
  [\pasp] {10.1086/670067}, \href
  {https://ui.adsabs.harvard.edu/abs/2013PASP..125..306F} {125, 306}

\bibitem[\protect\citeauthoryear{Gao, Springel  \& White}{Gao
  et~al.}{2005}]{Gao:2005ca}
Gao L.,  Springel V.,   White S. D.~M.,  2005, \mn@doi [Mon. Not. Roy. Astron.
  Soc.] {10.1111/j.1745-3933.2005.00084.x}, 363, L66

\bibitem[\protect\citeauthoryear{{Guo}, {White}, {Angulo}, {Henriques},
  {Lemson}, {Boylan-Kolchin}, {Thomas}  \& {Short}}{{Guo}
  et~al.}{2013}]{Guo2013}
{Guo} Q.,  {White} S.,  {Angulo} R.~E.,  {Henriques} B.,  {Lemson} G.,
  {Boylan-Kolchin} M.,  {Thomas} P.,   {Short} C.,  2013, \mn@doi [\mnras]
  {10.1093/mnras/sts115}, \href
  {http://adsabs.harvard.edu/abs/2013MNRAS.428.1351G} {428, 1351}

\bibitem[\protect\citeauthoryear{Hamilton}{Hamilton}{2000}]{Hamilton_2000}
Hamilton A. J.~S.,  2000, \mn@doi [Monthly Notices of the Royal Astronomical
  Society] {10.1046/j.1365-8711.2000.03071.x}, 312, 257–284

\bibitem[\protect\citeauthoryear{Joachimi et~al.,}{Joachimi
  et~al.}{2021}]{Joachimi_2021}
Joachimi B.,  et~al., 2021, \mn@doi [Astronomy & Astrophysics]
  {10.1051/0004-6361/202038831}, 646, A129

\bibitem[\protect\citeauthoryear{{Klypin}, {Trujillo-Gomez}  \&
  {Primack}}{{Klypin} et~al.}{2011}]{Klypin2011}
{Klypin} A.~A.,  {Trujillo-Gomez} S.,   {Primack} J.,  2011, \mn@doi [\apj]
  {10.1088/0004-637X/740/2/102}, \href
  {https://ui.adsabs.harvard.edu/abs/2011ApJ...740..102K} {740, 102}

\bibitem[\protect\citeauthoryear{{Kuijken} et~al.,}{{Kuijken}
  et~al.}{2019}]{Kuijken2019}
{Kuijken} K.,  et~al., 2019, \mn@doi [\aap] {10.1051/0004-6361/201834918},
  \href {https://ui.adsabs.harvard.edu/abs/2019A&A...625A...2K} {625, A2}

\bibitem[\protect\citeauthoryear{Mandelbaum, Tasitsiomi, Seljak, Kravtsov  \&
  Wechsler}{Mandelbaum et~al.}{2005}]{Mandelbaum:2004mq}
Mandelbaum R.,  Tasitsiomi A.,  Seljak U.,  Kravtsov A.~V.,   Wechsler R.~H.,
  2005, \mn@doi [Mon. Not. Roy. Astron. Soc.]
  {10.1111/j.1365-2966.2005.09417.x}, 362, 1451

\bibitem[\protect\citeauthoryear{{Mead}}{{Mead}}{2017}]{Mead2017}
{Mead} A.~J.,  2017, \mn@doi [\mnras] {10.1093/mnras/stw2312}, \href
  {http://adsabs.harvard.edu/abs/2017MNRAS.464.1282M} {464, 1282}

\bibitem[\protect\citeauthoryear{{Mead} \& {Peacock}}{{Mead} \&
  {Peacock}}{2014}]{Mead2014}
{Mead} A.~J.,  {Peacock} J.~A.,  2014, \mn@doi [\mnras] {10.1093/mnras/stu345},
  \href {http://adsabs.harvard.edu/abs/2014MNRAS.440.1233M} {440, 1233}

\bibitem[\protect\citeauthoryear{Mead \& Verde}{Mead \&
  Verde}{2021}]{Mead:2020qdk}
Mead A.~J.,  Verde L.,  2021, \mn@doi [Mon. Not. Roy. Astron. Soc.]
  {10.1093/mnras/stab748}, 503, 3095

\bibitem[\protect\citeauthoryear{Mead, Peacock, Heymans, Joudaki  \&
  Heavens}{Mead et~al.}{2015}]{Mead:2015yca}
Mead A.,  Peacock J.,  Heymans C.,  Joudaki S.,   Heavens A.,  2015, \mn@doi
  [Mon. Not. Roy. Astron. Soc.] {10.1093/mnras/stv2036}, 454, 1958

\bibitem[\protect\citeauthoryear{{Mead}, {Tr{\"o}ster}, {Heymans}, {Van
  Waerbeke}  \& {McCarthy}}{{Mead} et~al.}{2020}]{HMxCode}
{Mead} A.~J.,  {Tr{\"o}ster} T.,  {Heymans} C.,  {Van Waerbeke} L.,
  {McCarthy} I.~G.,  2020, \mn@doi [\aap] {10.1051/0004-6361/202038308}, \href
  {https://ui.adsabs.harvard.edu/abs/2020A&A...641A.130M} {641, A130}

\bibitem[\protect\citeauthoryear{Miyatake et~al.,}{Miyatake
  et~al.}{2020}]{Miyatake:2020uhg}
Miyatake H.,  et~al., 2020, arXiv e-prints, p. arXiv:2101.00113

\bibitem[\protect\citeauthoryear{{Miyatake} et~al.,}{{Miyatake}
  et~al.}{2021}]{miyatake/etal:2021}
{Miyatake} H.,  et~al., 2021, arXiv e-prints, \href
  {https://ui.adsabs.harvard.edu/abs/2021arXiv211102419M} {p. arXiv:2111.02419}

\bibitem[\protect\citeauthoryear{{More}, {van den Bosch}, {Cacciato}, {More},
  {Mo}  \& {Yang}}{{More} et~al.}{2013}]{More_2013}
{More} S.,  {van den Bosch} F.~C.,  {Cacciato} M.,  {More} A.,  {Mo} H.,
  {Yang} X.,  2013, \mn@doi [\mnras] {10.1093/mnras/sts697}, \href
  {https://ui.adsabs.harvard.edu/abs/2013MNRAS.430..747M} {430, 747}

\bibitem[\protect\citeauthoryear{{Navarro}, {Frenk}  \& {White}}{{Navarro}
  et~al.}{1996}]{1996NFWN}
{Navarro} J.~F.,  {Frenk} C.~S.,   {White} S. D.~M.,  1996, \apj, \href
  {https://ui.adsabs.harvard.edu/abs/1996ApJ...462..563N} {462, 563}

\bibitem[\protect\citeauthoryear{{Navarro}, {Frenk}  \& {White}}{{Navarro}
  et~al.}{1997}]{NFW_1997}
{Navarro} J.~F.,  {Frenk} C.~S.,   {White} S. D.~M.,  1997, \mn@doi [\apj]
  {10.1086/304888}, \href
  {https://ui.adsabs.harvard.edu/abs/1997ApJ...490..493N} {490, 493}

\bibitem[\protect\citeauthoryear{{Nishimichi} et~al.,}{{Nishimichi}
  et~al.}{2019}]{Nishimichi2019}
{Nishimichi} T.,  et~al., 2019, \mn@doi [\apj] {10.3847/1538-4357/ab3719},
  \href {https://ui.adsabs.harvard.edu/abs/2019ApJ...884...29N} {884, 29}

\bibitem[\protect\citeauthoryear{{Nishimichi} et~al.,}{{Nishimichi}
  et~al.}{2021}]{DarkEmulatorCode}
{Nishimichi} T.,  et~al., 2021, {DarkEmulator: Cosmological emulation code for
  halo clustering statistics} (\mn@eprint {ascl} {2103.009})

\bibitem[\protect\citeauthoryear{{Planck Collaboration} et~al.,}{{Planck
  Collaboration} et~al.}{2020}]{Planck2020}
{Planck Collaboration} et~al., 2020, \mn@doi [\aap]
  {10.1051/0004-6361/201833910}, \href
  {https://ui.adsabs.harvard.edu/abs/2020A&A...641A...6P} {641, A6}

\bibitem[\protect\citeauthoryear{{Prada}, {Klypin}, {Cuesta}, {Betancort-Rijo}
  \& {Primack}}{{Prada} et~al.}{2012}]{Prada2012}
{Prada} F.,  {Klypin} A.~A.,  {Cuesta} A.~J.,  {Betancort-Rijo} J.~E.,
  {Primack} J.,  2012, \mn@doi [\mnras] {10.1111/j.1365-2966.2012.21007.x},
  \href {https://ui.adsabs.harvard.edu/abs/2012MNRAS.423.3018P} {423, 3018}

\bibitem[\protect\citeauthoryear{{Riebe} et~al.,}{{Riebe}
  et~al.}{2013}]{Riebe2013AN....334..691R}
{Riebe} K.,  et~al., 2013, \mn@doi [Astronomische Nachrichten]
  {10.1002/asna.201211900}, \href
  {https://ui.adsabs.harvard.edu/abs/2013AN....334..691R} {334, 691}

\bibitem[\protect\citeauthoryear{{Ruiz}, {Padilla}, {Dom{\'{\i}}nguez}  \&
  {Cora}}{{Ruiz} et~al.}{2011}]{Ruiz2011}
{Ruiz} A.~N.,  {Padilla} N.~D.,  {Dom{\'{\i}}nguez} M.~J.,   {Cora} S.~A.,
  2011, \mn@doi [\mnras] {10.1111/j.1365-2966.2011.19635.x}, \href
  {http://adsabs.harvard.edu/abs/2011MNRAS.418.2422R} {418, 2422}

\bibitem[\protect\citeauthoryear{Seljak, Hamaus  \& Desjacques}{Seljak
  et~al.}{2009}]{Seljak_2009}
Seljak U.,  Hamaus N.,   Desjacques V.,  2009, \mn@doi [Physical Review
  Letters] {10.1103/physrevlett.103.091303}, 103

\bibitem[\protect\citeauthoryear{{Sheldon} et~al.,}{{Sheldon}
  et~al.}{2004}]{Sheldon2004}
{Sheldon} E.~S.,  et~al., 2004, \mn@doi [\aj] {10.1086/383293}, \href
  {https://ui.adsabs.harvard.edu/abs/2004AJ....127.2544S} {127, 2544}

\bibitem[\protect\citeauthoryear{Sheth \& Tormen}{Sheth \&
  Tormen}{1999}]{Sheth1999}
Sheth R.~K.,  Tormen G.,  1999, \mn@doi [\mnras]
  {10.1046/j.1365-8711.1999.02692.x}, 308, 119

\bibitem[\protect\citeauthoryear{Smith et~al.,}{Smith
  et~al.}{2003}]{Smith:2002dz}
Smith R.~E.,  et~al., 2003, \mn@doi [Mon. Not. Roy. Astron. Soc.]
  {10.1046/j.1365-8711.2003.06503.x}, 341, 1311

\bibitem[\protect\citeauthoryear{Takahashi, Sato, Nishimichi, Taruya  \&
  Oguri}{Takahashi et~al.}{2012}]{Takahashi:2012em}
Takahashi R.,  Sato M.,  Nishimichi T.,  Taruya A.,   Oguri M.,  2012, \mn@doi
  [Astrophys. J.] {10.1088/0004-637X/761/2/152}, 761, 152

\bibitem[\protect\citeauthoryear{Tinker, Weinberg, Zheng  \& Zehavi}{Tinker
  et~al.}{2005}]{Tinker:2004gf}
Tinker J.~L.,  Weinberg D.~H.,  Zheng Z.,   Zehavi I.,  2005, \mn@doi
  [Astrophys. J.] {10.1086/432084}, 631, 41

\bibitem[\protect\citeauthoryear{{Tinker}, {Robertson}, {Kravtsov}, {Klypin},
  {Warren}, {Yepes}  \& {Gottl{\"o}ber}}{{Tinker} et~al.}{2010a}]{Tinker2010}
{Tinker} J.~L.,  {Robertson} B.~E.,  {Kravtsov} A.~V.,  {Klypin} A.,  {Warren}
  M.~S.,  {Yepes} G.,   {Gottl{\"o}ber} S.,  2010a, \mn@doi [\apj]
  {10.1088/0004-637X/724/2/878}, \href
  {http://adsabs.harvard.edu/abs/2010ApJ...724..878T} {724, 878}

\bibitem[\protect\citeauthoryear{{Tinker}, {Robertson}, {Kravtsov}, {Klypin},
  {Warren}, {Yepes}  \& {Gottl{\"o}ber}}{{Tinker} et~al.}{2010b}]{2010tinker}
{Tinker} J.~L.,  {Robertson} B.~E.,  {Kravtsov} A.~V.,  {Klypin} A.,  {Warren}
  M.~S.,  {Yepes} G.,   {Gottl{\"o}ber} S.,  2010b, \mn@doi [\apj]
  {10.1088/0004-637X/724/2/878}, \href
  {https://ui.adsabs.harvard.edu/abs/2010ApJ...724..878T} {724, 878}

\bibitem[\protect\citeauthoryear{{Tr{\"o}ster} et~al.,}{{Tr{\"o}ster}
  et~al.}{2021}]{Troester_2021}
{Tr{\"o}ster} T.,  et~al., 2021, arXiv e-prints, \href
  {https://ui.adsabs.harvard.edu/abs/2021arXiv210904458T} {p. arXiv:2109.04458}

\bibitem[\protect\citeauthoryear{\VAN{Bosch}{Van}{van}~{den Bosch}, {More},
  {Cacciato}, {Mo}  \& {Yang}}{\VAN{Bosch}{Van}{van}~{den Bosch}
  et~al.}{2013}]{2013BoschCacciato}
\VAN{Bosch}{Van}{van}~{den Bosch} F.~C.,  {More} S.,  {Cacciato} M.,  {Mo} H.,
   {Yang} X.,  2013, \mnras, \href
  {https://ui.adsabs.harvard.edu/abs/2013MNRAS.430..725V} {430, 725}

\bibitem[\protect\citeauthoryear{\VAN{Uitert}{Van}{van}~Uitert
  et~al.,}{\VAN{Uitert}{Van}{van}~Uitert et~al.}{2016}]{vanUitert2016}
\VAN{Uitert}{Van}{van}~Uitert E.,  et~al., 2016, \mn@doi [\mnras]
  {10.1093/mnras/stw747}, \href
  {https://ui.adsabs.harvard.edu/abs/2016MNRAS.459.3251V} {459, 3251}

\bibitem[\protect\citeauthoryear{{Wang} et~al.,}{{Wang}
  et~al.}{2013}]{Wang2013}
{Wang} L.,  et~al., 2013, \mn@doi [\mnras] {10.1093/mnras/stt190}, \href
  {https://ui.adsabs.harvard.edu/abs/2013MNRAS.431..648W} {431, 648}

\bibitem[\protect\citeauthoryear{Wechsler, Zentner, Bullock  \&
  Kravtsov}{Wechsler et~al.}{2006}]{Wechsler:2005gb}
Wechsler R.~H.,  Zentner A.~R.,  Bullock J.~S.,   Kravtsov A.~V.,  2006,
  \mn@doi [Astrophys. J.] {10.1086/507120}, 652, 71

\bibitem[\protect\citeauthoryear{Yang, Mo  \& Bosch}{Yang
  et~al.}{2008}]{Yang:2007pg}
Yang X.,  Mo H.~J.,   Bosch F. C. v.~d.,  2008, \mn@doi [Astrophys. J.]
  {10.1086/528954}, 676, 248

\bibitem[\protect\citeauthoryear{Zacharegkas et~al.}{Zacharegkas
  et~al.}{2022}]{DES:2021olg}
Zacharegkas G.,  et~al., 2022, \mn@doi [Mon. Not. Roy. Astron. Soc.]
  {10.1093/mnras/stab3155}, 509, 3119

\bibitem[\protect\citeauthoryear{{Zel'dovich}}{{Zel'dovich}}{1970}]{Zeldovich1970}
{Zel'dovich} Y.~B.,  1970, AAP, \href
  {http://adsabs.harvard.edu/abs/1970A%26A.....5...84Z} {5, 84}

\bibitem[\protect\citeauthoryear{Zennaro, Angulo, Aric\`o, Contreras  \&
  Pellejero-Ib\'a\~nez}{Zennaro et~al.}{2019}]{Zennaro2019}
Zennaro M.,  Angulo R.~E.,  Aric\`o G.,  Contreras S.,   Pellejero-Ib\'a\~nez
  M.,  2019, \mn@doi [Mon. Not. Roy. Astron. Soc.] {10.1093/mnras/stz2612},
  489, 5938

\bibitem[\protect\citeauthoryear{{Zennaro}, {Angulo},
  {Pellejero-Ib{\'a}{\~n}ez}, {St{\"u}cker}, {Contreras}  \&
  {Aric{\`o}}}{{Zennaro} et~al.}{2021}]{Zennaro2021}
{Zennaro} M.,  {Angulo} R.~E.,  {Pellejero-Ib{\'a}{\~n}ez} M.,  {St{\"u}cker}
  J.,  {Contreras} S.,   {Aric{\`o}} G.,  2021, arXiv e-prints, \href
  {https://ui.adsabs.harvard.edu/abs/2021arXiv210112187Z} {p. arXiv:2101.12187}

\bibitem[\protect\citeauthoryear{Zheng et~al.,}{Zheng
  et~al.}{2005}]{Zheng:2004id}
Zheng Z.,  et~al., 2005, \mn@doi [Astrophys. J.] {10.1086/466510}, 633, 791

\makeatother
\end{thebibliography}



\appendix
\section{Dark Quest and the \textsc{Dark Emulator}}
\label{app:DarkEmu}
In this Appendix we review the Dark Quest $N$-body simulations used to calibrate the beyond-linear halo bias quantity $\Bnl$, referring the reader to \citet{Nishimichi2019,DarkEmulatorCode} for full details.  Dark Quest explores $100$ sets of $w$CDM cosmological parameters selected in six-dimensional space using a latin hypercube design. At each cosmological parameter set, a high-resolution and a low-resolution simulation are performed. The former (latter) covers comoving cubes with the side length of $1\,h^{-1}\mathrm{Gpc}$ ($2\,h^{-1}\mathrm{Gpc}$), while the number of simulation particles is fixed to $2,048^3$. The \textsc{Rockstar} finder \citep{Rockstar} is applied to identify dark matter halos, and they are analyzed after subhalos are removed. The halo and matter two-point correlation functions (both auto and cross), as well as the halo mass function, are tabulated at various halo masses\footnote{The cumulative halo number density was used as a proxy of the mass in the actual emulator implementation. They can be converted to each other using the mass function emulator.} and redshifts. The \textsc{DarkEmulator}\footnote{\textsc{DarkEmulator}:\url{https://github.com/DarkQuestCosmology/dark_emulator_public}} regressor utilises Gaussian Processes and a weighted Principal Component Analysis to make predictions from Dark Quest at any set of cosmological parameters within the support range of the training simulations (\citealt{Nishimichi2019}, equation 25).  It makes use of the FFTLog algorithm \citep{Hamilton_2000} to quickly move from configuration space to Fourier space. The \citet{Zheng:2004id} halo occupation distribution (HOD) model is also implemented to make predictions for galaxy statistics \citep{Miyatake:2020uhg}.

To determine $\Bnl$, equation~\ref{eq:bnl hh}, we use the \textsc{DarkEmulator} to predict the quantities of the linear bias $\hat{b}(M,z)$ and the halo auto power spectrum $\hat{P}_\mathrm{hh}(M_1,M_2,k,z)$.    Direct measurements of the Dark Quest real-space halo-halo two-point correlation function, $\xi_\mathrm{hh}$, are smoothly connected to an analytical prescription on large scales to mitigate the impact of sample variance noise.  This is found to be significant even with the $(2\,h^{-1}\mathrm{Gpc})^3$ volume of the low-resolution simulation suite. The scale to switch to the analytical perscription is set to $60\,h^{-1}\mathrm{Mpc}$, with the large-scale signal taking the form
\begin{equation}
\xi_\mathrm{hh}(r,z,M_1,M_2) = \mathrm{IFT}\left[\Gamma_\mathrm{h}(k,z,M_1)\Gamma_\mathrm{h}(k,z,M_2)P^\mathrm{lin}_{\delta\delta}(k,z)\right].
\label{eq:large_scale_limit}
\end{equation}
Here IFT stands for an inverse Fourier Transform and the function $\Gamma_\mathrm{h}$ is the propagator defined by
\begin{equation}
\Gamma_\mathrm{h}(k,z,M) = \dfrac{P_\mathrm{h\delta}^\mathrm{lin}(k,z,M)}{P_\mathrm{\delta\delta}^\mathrm{lin}(k,z)},
\end{equation}
with $P_\mathrm{h\delta}^\mathrm{lin}$ being the \citet{Crocce_2006} cross spectrum between the halo density field and the linear matter density field. This function exhibits a simple, near Gaussian, damping behavior towards high-$k$, which describes the damping of the bump in the correlation function originating from baryon acoustic oscillations. At the other end, the low-$k$ limit of $\Gamma_\mathrm{h}$ corresponds to the linear bias factor. Our $\hat{P}_\mathrm{hh}$ used to estimate $\beta_\mathrm{NL}$ is the Fourier Transform of the $\xi_\mathrm{hh}$ function. The \textsc{DarkEmulator} module for $\Gamma_\mathrm{h}$ is used to evaluate $b(M)$ and is consistently calibrated against the low-resolution simulations, which have less statistical uncertainties.

It is worth noting that there is no guarantee that the function $\beta^{\mathrm{NL}}$ evaluated this way approaches zero in the low-$k$ limit (see Figure \ref{fig:daremu_comp}). This is because of the mixture of scales in the Fourier Transform. Our halo power spectrum does not necessarily approach to $\Gamma_\mathrm{h}(k,z,M_1)\Gamma_\mathrm{h}(k,z,M_2)P^\mathrm{lin}_{\delta\delta}(k,z)$ at low-$k$ despite the use of the prescription in Eq.~(\ref{eq:large_scale_limit}). Indeed, effects, such as the halo-exclusion effect, which are confined to small scales in configuration space, are known to contribute to the low-$k$ part of the power spectrum, leading to non-Poissonian shot noise \citep{Seljak_2009,Baldauf_2013}. \textsc{DarkEmulator} automatically takes account of these physical effects in its predictions.

To reduce computation time we determine $\Bnl$ from the \textsc{DarkEmulator} on a regular grid of $k$, $M_1$, $M_2$ and $z$, then construct an interpolator. For $k$ we take 50 points between $10^{-2}$ and $10^{1.5} \ \mathrm{h/Mpc}$, for $M_1$ and $M_2$, 5 points between $10^{12}$ and $10^{14} \ \mathrm{M_\odot/h}$, and for $z$, 5 points between 0.0 and 0.5. We construct an interpolator for $\Bnl$ using linear interpolation, extrapolating $\Bnl$ outside of the domain. This interpolation process will be optimised in future work.

\section{Rescaling halo bias} \label{sec:rescaling}

In this paper we have demonstrated the importance of including beyond-linear halo bias and halo exclusion, both incorporated within $\Bnl$, for calculations that involve the halo model. In this appendix we consider the cosmology dependence of $\Bnl$ and present a rescaling technique that is able to predict the cosmology dependence with reasonable success.

First, recall that $\Bnl(M_1, M_2, k)$ is really a proxy for $P_\mathrm{hh}(M_1, M_2, k)$. \cite{Mead:2020qdk} advocated using $\Bnl$, the ratio of the halo--halo spectrum to the linear spectrum, rather than the power spectrum directly because this ratio will cancel some of the cosmology dependence intrinsic to $P_\mathrm{hh}$, for example, the large-scale dependence on $\sigma_8$. In fact, because $\Bnl$ is designed to be zero at large scales, independent of cosmology, already means that a significant amount of the cosmology dependence is absorbed in its initial definition.

To address the further cosmology dependence of $\Bnl$ we utilize the `rescaling' technique of \citeauthor{Angulo2010} (\citeyear{Angulo2010}; AW10) to map the function between different cosmologies. AW10 proposed a redefinition (or rescaling) of length and time units (which together imply a mass-unit rescaling) of a cosmological \nbody simulation, chosen such that the halo-mass function that would be inferred from the rescaled simulation closely matched that in a desired `target' simulation, with different cosmology. After length and time rescaling, the \cite{Zeldovich1970} approximation can be used to adjust the large-scale displacement field of the particle distribution to account for residual differences in the linear clustering between the rescaled and target cosmology. In principle, the rescaling can be applied multiple times from the same original simulation, and so a single simulation can be used to model properties of multiple different cosmologies. AW10 demonstrated that  clustering statistics for matter and for haloes measured from rescaled simulations compared well to those from proper simulations of the target cosmology. The algorithm has been further tested and developed: \cite{Ruiz2011} and \cite{Mead2014} demonstrated that rescaling can be applied to haloes directly, without the need to go via the simulated matter distribution. \cite{Guo2013} showed the properties of galaxy distributions were robustly reproduced under rescaling. More recently, the algorithm has been extended to massive-neutrino cosmologies by \cite{Zennaro2019}, and to baryonic physics by \cite{Giovanni2020}. Rescaling has been used recently to greatly reduce the computational burden of building cosmological emulators \citep{Contreras2020, Zennaro2021}.

We test the rescaling approach using the \DQ emulator of \cite{Nishimichi2019}, which can be used to emulate the halo--halo power spectra over a range of masses and cosmologies and therefore to construct $\Bnl$. The stated accuracy of \DQ for halo--halo power spectra is $4$ per cent, which sets a limit to how well we can use the emulator to probe the rescaling technique. Given the existence of \DQ, it is clearly not necessary to perform this rescaling, but in the future we envisage cosmological analyses wanting to explore parameter space beyond the \DQ hypercube and therefore some means to extrapolate results from the emulator become essential.

While the usual AW10 algorithm is applied directly to particle or halo data from \nbody simulations, there is no reason not to apply the algorithm to a summary statistic, such as $\Bnl$, that has already been measured from a simulation; although a disadvantage of doing this is that the final `displacement field' step of the algorithm cannot be applied. Following AW10, we decide on a `target' cosmology at redshift $z'$ and we attempt to match that cosmology by rescaling a `fiducial' cosmology by evaluating quantities of interest in that cosmology at a redshift $z$ and then by rescaling length units by dimensionless parameter $s$, such that:
\begin{equation}
\frac{R'}{h'^{-1}\mathrm{Mpc}} = \frac{sR}{h^{-1}\mathrm{Mpc}}\ .
\label{eq:rescaling_R}
\end{equation}
Mass conservation implies that this length rescaling simultaneously implies a mass rescaling:
\begin{equation}
\frac{M'}{h'^{-1}M_\odot} = \frac{\Omega^{'}_\mathrm{m}}{\Om}\frac{s^3M}{h^{-1}M_\odot}
= \frac{s_\mathrm{m}M}{h^{-1}M_\odot}\ .
\label{eq:rescaling_M}
\end{equation}
Note carefully the factors of $h$ and $h'$ in the units that arise in equations~(\ref{eq:rescaling_R}) and (\ref{eq:rescaling_M}), which appear because of the standard convention to use factors of $h$ in some cosmological units. Primed quantities are in the target cosmology while unprimed are those in the fiducial cosmology. We can calculate the variance in the density field when smoothed on comoving
scale $R$, $\sigma(R)$, in any cosmology as it only relies on linear theory,
\begin{equation}
  \sigma^2(R) =\int_0^\infty \Delta^2_\mathrm{lin}(k)T^2(kR)\,\mathrm{d} \ln k\ ,
  \label{eq:sigmaR}
  \end{equation}
$T(x)$ is the spherical Fourier transform of a top-hat window function. Most prescriptions for the halo mass function \citep[\eg][]{Sheth1999, Tinker2010, Despali2016} are parameterised in terms of $\sigma(R)$, which has been shown to be the quantity of primary interest for halo formation \citep[][]{Bond1991}. We therefore use $\sigma(R)$ to find a match between the fiducial and the target cosmologies by minimising the `cost function', 
\begin{equation}
  \delta^2(s,z)=\frac{1}{\ln(R'_2/R'_1)}\int_{R'_1}^{R'_2}\frac{\mathrm{d}R'}{R'}
  \left[1-\frac{\sigma(s^{-1}R', z)}{\sigma'(R',z')}\right]^2\ ,
  \label{eq:rescaling_cost}
\end{equation}
which is equivalent to the ratio of $\sigma(R)$ functions across a logarithmic range in $R$. Note that choosing $s$ and $z$ according to equation~(\ref{eq:rescaling_cost}) usually results in the linear spectra also being closely matched because of the close relationship between the two as evidenced by equation~(\ref{eq:sigmaR}). The range between $R'_1$ and $R'_2$ is chosen to correspond to the Lagrangian radii of haloes in the desired target sample.

\begin{table}
\caption{Rescaling parameters for the cosmologies we consider in this appendix. In each case we rescale the fiducial (central) \DQ cosmology, with parameters \{$\oc=0.120$; $\ob=0.0223$; $\Om=0.316$; $\ns=0.965$; $\As=\sform{2.21}{-9}$; $w=-1$\} to match the target cosmology at $z'=0.5$ with the `deviant' parameter noted in the first column (only one parameter is varied at a time). The columns of the table should therefore be thought of as the redshift $z$ and size scaling $s$ that need to be applied to the fiducial cosmology to match the target. Note that the rescaling parameters are most severe in $z$ for the $\As$ scaling, because this parameter has the biggest effect on the power spectrum of those within the \DQ hypercube.}
\begin{center}
\begin{tabular}{c c c c c}
\hline
Target deviant parameter & $s$ & $z$ & $s_\mathrm{m}$ & $h'/h$ \\
\hline
$\oc=0.1114$ & $1.049$ & $0.673$ & $1.153$ & $0.970$ \\
$\oc=0.1282$ & $0.957$ & $0.346$ & $0.877$ & $1.029$ \\
$\ob=0.0215$ & $0.995$ & $0.482$ & $0.984$ & $0.997$ \\
$\ob=0.0230$ & $1.005$ & $0.518$ & $1.016$ & $1.002$ \\
$\Ow=0.5886$ & $0.876$ & $0.444$ & $0.876$ & $0.876$ \\
$\Ow=0.7802$ & $1.198$ & $0.592$ & $1.198$ & $1.198$ \\
$\As=\sform{1.4308}{-9}$ & $1.000$ & $0.954$ & $1.000$ & $1.000$ \\
$\As=\sform{3.4027}{-9}$ & $1.000$ & $0.087$ & $1.000$ & $1.000$ \\
$\ns=0.9307$ & $1.080$ & $0.632$ & $1.261$ & $1.000$ \\
$\ns=0.9983$ & $0.927$ & $0.372$ & $0.796$ & $1.000$ \\
$w=-1.14$ & $1.000$ & $0.454$ & $1.000$ & $1.000$ \\
$w=-0.86$ & $1.000$ & $0.566$ & $1.000$ & $1.000$ \\
\hline
\end{tabular}
\end{center}
\label{tab:rescaling}
\end{table}

In our case, we choose $\log_{10}(M'_1/h'^{-1}M_\odot)=12.5$ and $\log_{10}(M'_2/h'^{-1}M_\odot)=15$ for our halo-mass range, which corresponds to the range of haloes probed by \DQ. We choose our `fiducial' cosmology to be at the centre of the \DQ parameter hypercube (parameters in the caption of Table~\ref{tab:rescaling}) and we test how well the rescaling algorithm allows us to match $\Bnl$ at $z'=0.5$ for different cosmologies around the parameter hypercube. The values of the rescaling parameters $s$, $z$, and $s_\mathrm{m}$ for each cosmology are given in Table~\ref{tab:rescaling}. For most cosmologies, the rescaling represents only a small change ($s$ is close to unity and $z$ is close to $z'=0.5$); the exception is scaling in $\As$, which requires comparatively large changes in $z$. This is due to the comparatively large range of $\As$ spanned by the emulator (a factor of $\sim3.5$ in $\As$, corresponding to a factor of $\sim1.8$ in $\sigma_8$). Note that for cosmologies that change only $\As$ and $w$, $s=1$ is required, which is because the linear theory power spectra for these models have identical shapes and are only offset in amplitude, which can always be mapped to a different $z$ for scale-independent linear growth. Note also that for some cosmologies (\eg high/low $\Ow$) the required value of $s$ is identical to $h'/h$, which indicates that the linear spectrum shapes are identical but with pure horizontal and vertical offsets, with the horizontal offset purely a function of our decision to use $\Mpc$ units, rather than pure $\mathrm{Mpc}$. For all cosmologies considered in this appendix, choosing $s$ and $z$ via equation~(\ref{eq:rescaling_cost}) results in near perfect matches to the $\sigma(M)$ function of the target cosmology, with residuals well below the per-cent level across all relevant scales.

Once $s$, $s_\mathrm{m}$ and $z$ have been computed via the minimization of equation~(\ref{eq:rescaling_cost}), we evaluate the $\Bnl$ function in the original cosmology and compare it to that in the target cosmology: \ie comparing rescaled
\begin{equation}
\beta^\mathrm{NL}_\mathrm{res}=\beta^\mathrm{NL}(M_1=M'_1/s_\mathrm{m}, M_2=M'_2/s_\mathrm{m}, k=sk', z=z)\ ,
\end{equation}
with $\Bnl$ evaluated in the fiducial cosmology, to target $\beta^\mathrm{NL}_\mathrm{tgt}=\beta^\mathrm{'NL}(M'_1, M'_2, k', z')$. We also show the `standard' comparison, without rescaling, where we simply evaluate the fiducial $\Bnl$ at the target masses, wavenumbers and redshift, \ie 
\begin{equation}
\beta^\mathrm{NL}_\mathrm{std}=\beta^\mathrm{NL}(M_1=M'_1, M_2=M'_2, k=k', z=z')\ .
\end{equation}

At fixed redshift, $\Bnl$ is a function of three variables, which makes it unwieldy to plot a comparison of rescaled and target versions. We therefore create a one-dimensional summary statistic
\begin{equation}
\begin{split}
&\sigma^2_\beta(k')=\frac{1}{\ln(M'_2/M'_1)^2}
\int_{M'_1}^{M'_2}\int_{M'_1}^{M'_2}
\,\mathrm{d}\ln M'_1\,\mathrm{d}\ln M'_2 \\
&\quad\times\left[\beta^\mathrm{NL}(M'_1/s_\mathrm{m}, M'_2/s_\mathrm{m}, sk')-\beta^{'\mathrm{NL}}(M'_1, M'_2, k')\right]^2
\ ,
\end{split}
\label{eq:sigma_k}
\end{equation}
to asses the performance of the rescaling, which corresponds to a mean difference over logarithmic ranges in both halo-mass variables. We also considered weighting the above integral by factors of the halo-mass function, but decided against this because different calculations are sensitive to different halo mass ranges, and a mass-function weighting strongly boosts the contribution from lower halo masses. $\Bnl$ itself has a roughly similar shape and amplitude for all halo-mass arguments, so equation~(\ref{eq:sigma_k}) has the advantage of roughly evenly weighting in log halo mass.

\begin{figure*}
\centering
\includegraphics[width=2\columnwidth]{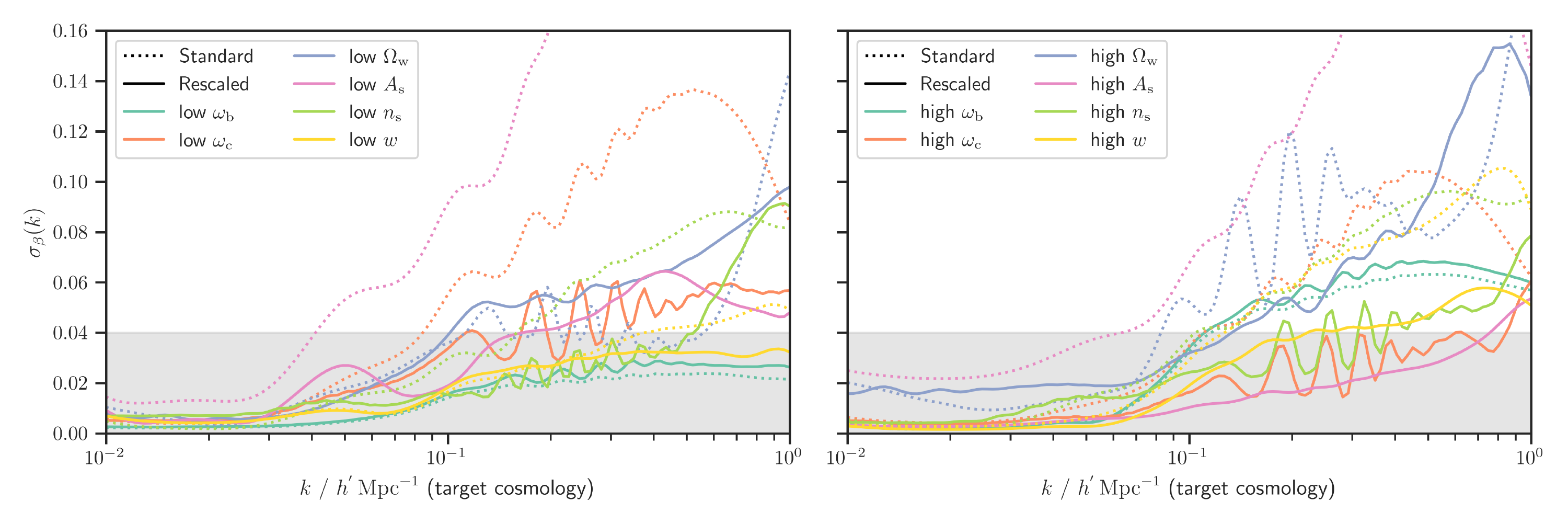}
\caption{Performance of the rescaling algorithm for the case when $\Bnl$ from the cosmology at the centre of the \DQ parameter space is rescaled to match the cosmology denoted in the plot legend. The left-hand panel shows the performance when attempting to match lower values of the cosmological parameters whereas the right-hand panel show the performance for higher values of the cosmological parameters. Dotted curves show pre-rescaling (from which the intrinsic cosmology dependence of $\Bnl$ can be inferred), while solid curves show post rescaling. The grey region denotes the quoted $4$ per-cent error for the halo--halo power spectrum from \DQ. It is clear that performing the rescaling on $\Bnl$ provides a better match to the target data, with errors comparable to the emulator performance across most relevant scales.}
\label{fig:rescaling}
\end{figure*}

\begin{figure}
  \centering
  \includegraphics[width=1\columnwidth]{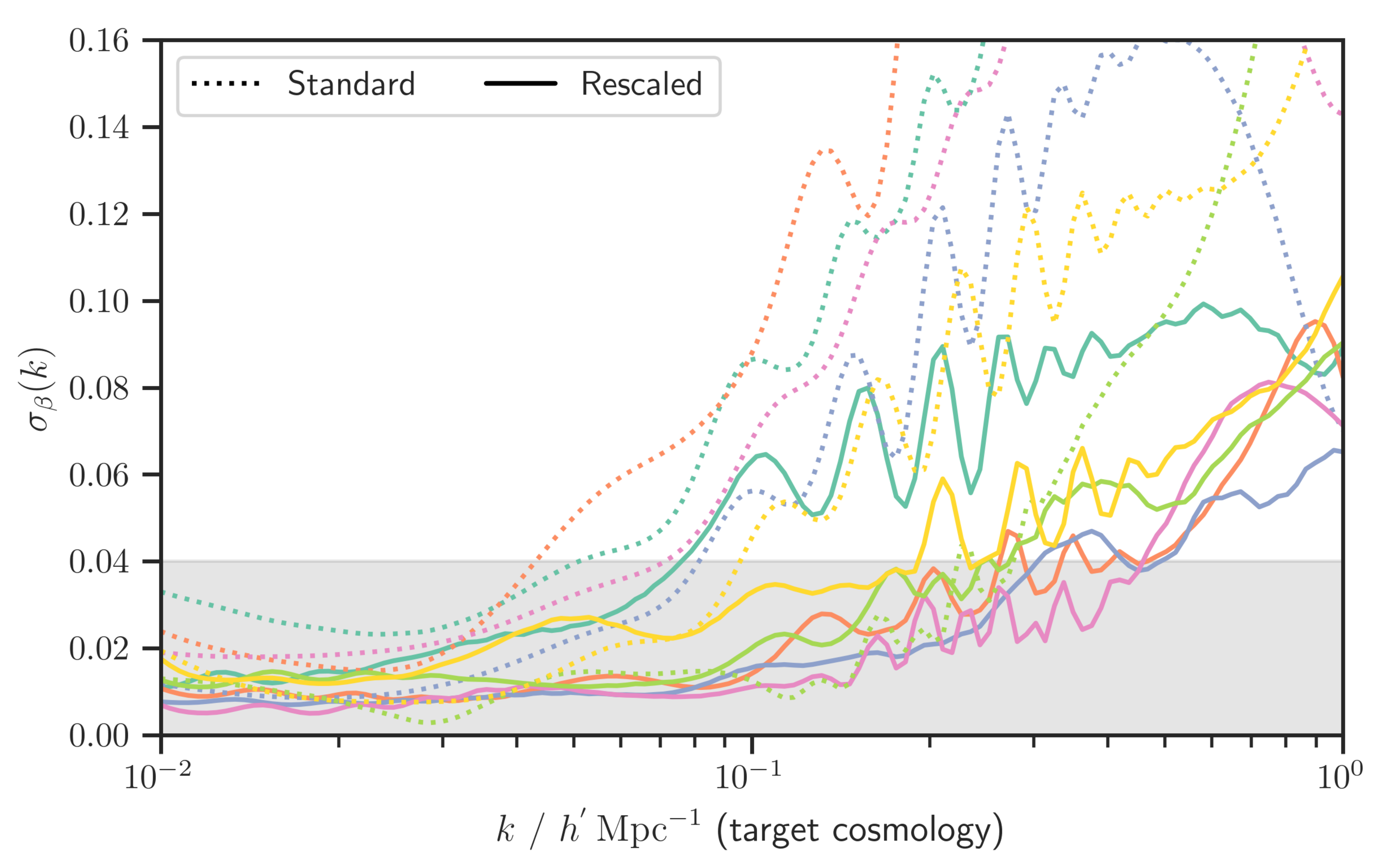}
  \caption{Similar to Fig.~\ref{fig:rescaling}, but each coloured line represents a `random' cosmology drawn uniformly from within the \DQ parameter hypercube. Note that there is no correspondence between the cosmological models shown in the left and right panels, even when they share a colour. In all cases, and for all scales shown here, the rescaling improves the correspondence with the target cosmology. The error mostly stays below $10$ per cent.}
  \label{fig:rescaling_random}
  \end{figure}

In Fig.~\ref{fig:rescaling} we show the performance of the rescaling algorithm via the summary statistic given in equation~(\ref{eq:sigma_k}). In almost all cases, and over almost all scales, the rescaling improves the match of the original to the target $\Bnl$. The exceptions are for the changing $\ob$, where it has a tiny, detrimental effect, and for the low $\oc$ model, where rescaling also degrades the match. The shaded-grey area shows the quoted $4$ per cent \DQ error; it would be unrealistic to expect our results to be better than this limit. In all cases, the error stays below $6$ per cent for $k<0.3\iMpc$. The error remains at this level for $k<1\iMpc$ for all cosmologies except those that change $\Ow$, where the error can reach a maximum $16$ per cent. In Fig.~\ref{fig:rescaling_random} we instead show results for $12$ random cosmologies, drawn uniformly from the \DQ hypercube. Results are similar, but slightly degraded compared to Fig.~\ref{fig:rescaling} where we vary only a single cosmological parameter for each target model. However, the error mainly stays below $10$ per cent in all cases shown. Note that we envisage $\Bnl$ being applied as a \emph{correction} to standard halo model calculations, where it represents an at-most $30$ per-cent correction; therefore a $10$ per-cent error in $\Bnl$ translates to a $\sim 3$ per-cent error in the eventual halo model. We consider these results highly encouraging for the idea of using rescaling to estimate $\Bnl$ outside the \DQ parameter space.

In future, we could consider including minimizing the difference in growth function in our choice of $s$ and $z$ (together with the standard equation~\ref{eq:rescaling_cost}) using the extended method proposed by \citet{Angulo2015}. They refined the choice of rescaling parameters to consider the \emph{historical} structure formation in the target cosmology (as well as the mass function), and demonstrated that doing this improves the match for the rescaled halo concentration--mass relation, which itself has been shown to be dependent on the halo formation history \citep[\eg][]{Bullock2001}. It may be that the beyond-linear clustering of haloes is sensitive to the structure-formation history, rather than just the present day linear spectrum shape and amplitude \citep[\eg][]{Mead2017}. It would also be interesting to measure $\Bnl$ in rescaled simulations directly, which has the advantage that the displacement-field step can be applied to the halo catalogue pre measurement. Since $\Bnl$ exclusively contains beyond-linear physics this would be a test of the generality of the displacement-field step beyond linear scales. Halo masses in \DQ are defined using the $\times 200$ background density spherical-overdensity criterion, but various authors \citep[\eg][]{Despali2016, Mead2017} have suggested that a cosmology-dependent virial overdensity criterion, informed via the spherical-collapse model, may better capture the cosmology dependence. Indeed, the results in Fig.~\ref{fig:rescaling} were least good for changing $\Ow$, which has the largest effect on spherical-collapse calculations. Clearly $\Bnl$ is a function of the halo-identification technique (both overdensity and so-called percolation) and so we are not in a position to test the performance of rescaling with virial-defined haloes, but this would be an intriguing direction for future work.


\bsp	
\label{lastpage}
\end{document}